\documentclass[preprint,12pt]{elsarticle}

\usepackage[margin=1in]{geometry}
\usepackage[x11names]{xcolor}
\usepackage{graphicx}
\usepackage{amsmath, amsthm, mathtools, amssymb, color, float, eqnarray, booktabs, makecell, soul, mdframed, multirow}
\usepackage[hyphens]{url}
\usepackage[justification=centering]{caption}
\usepackage{todonotes}
\usepackage{comment}
\usepackage{bm}
\usepackage{ulem}
\usepackage[most]{tcolorbox}
\usepackage{array}
\usepackage{tabularx}
\usepackage{amsfonts} 
\usepackage{breqn}
\usepackage[colorlinks=true, linkcolor=blue, citecolor=blue, urlcolor=blue]{hyperref}
\usepackage{cleveref}
\usepackage{subcaption} 
\usepackage[utf8]{inputenc}
\usepackage{csquotes}
\usepackage[T1]{fontenc}     
\usepackage{enumitem}
\usepackage{bm}
\usepackage{multirow}
\usepackage{pdflscape}
\usepackage{float}
\usepackage{caption}
\usepackage{xcolor}
\definecolor{darkred}{rgb}{0.9,0.1,0.1}
\usepackage{booktabs} 
\usepackage[ruled,vlined]{algorithm2e}
\hypersetup{
 colorlinks,
 citecolor=black,
 filecolor=black,
 linkcolor=black,
 urlcolor=black
}
\usepackage{amssymb}

\journal{arXiv}

\begin{document}

\begin{frontmatter}


\title{
Representation Meets Optimization: Training PINNs and PIKANs for Gray-Box Discovery in Systems Pharmacology}

\author[inst1]{Nazanin Ahmadi Daryakenari}
\author[inst2]{Khemraj Shukla}
\author[inst2]{George Em Karniadakis}
\affiliation[inst1]{organization={Center for Biomedical Engineering, Brown University},
  city={Providence},
  postcode={02912}, 
  state={RI},
  country={USA}}
\affiliation[inst2]{organization={Division of Applied Mathematics, Brown University},
  city={Providence},
  postcode={02912}, 
  state={RI},
  country={USA}}

\begin{abstract}
Physics-Informed Kolmogorov–Arnold Networks (PIKANs) have been gaining attention as an effective counterpart to the original multilayer perceptron-based Physics-Informed Neural Networks (PINNs).
Both representation models can address inverse problems and facilitate gray-box system identification. However, a comprehensive understanding of their performance in terms of accuracy and speed remains underexplored. In particular, we introduce a modified PIKAN architecture, tanh-cPIKAN, which is based on Chebyshev polynomials for parametrization of the univariate functions with an extra nonlinearity for enhanced performance.
We then present a systematic investigation of how the choices of optimizer, representation, and training configuration influence the performance of PINNs and PIKANs in the context of systems pharmacology modeling. We benchmark a wide range of optimizers using simple but representative pharmacokinetic and pharmacodynamic models. We use the new \texttt{Optax} library \cite{deepmind2020jax} but also a new class of self-scaled optimizers developed in \texttt{Optimistix} library \cite{optimistix} to identify the most effective combinations for learning gray-boxes  under ill-posed, non-unique, and data-sparse conditions. 
We examine the influence of model architecture (MLP vs. KAN), numerical precision (single vs. double), the need for warm-up phases for second-order methods, and sensitivity to the initial learning rate. We also assess the optimizer scalability for larger models and analyze the trade-offs introduced by JAX in terms of computational efficiency and numerical accuracy. Using two representative system pharmacology examples, a pharmacokinetics model and a chemotherapy drug response model, we offer practical guidance on selecting optimizers and representation models/architectures for robust and efficient gray-box discovery. Our findings provide actionable insights for improving the training of physics-informed networks in systems pharmacology, systems biology, and beyond.

\end{abstract}



\end{frontmatter}

\section{Introduction}

Inverse problems are omni-present across scientific domains, enabling the inference of hidden parameters and dynamic system behaviors from empirical observations. Such problems are especially significant in biology, physics, and pharmacology, where direct measurements are often impractical, and systems exhibit stiff, nonlinear dynamics. Traditional methods such as Sparse Identification of Nonlinear Dynamics (SINDy) \cite{brunton2016discovering}, which build on sparse regression \cite{tibshirani1996regression} and compressed sensing \cite{donoho2006compressed}, have demonstrated success in extracting interpretable equations from data. However, their reliance on predefined function libraries limits their expressiveness. Recent extensions like ADAM-SINDy \cite{adamSINDy2024} address this by jointly optimizing coefficients and nonlinear parameters via first-order methods such as Adam, thereby enhancing robustness.

Other symbolic regression methods—including AI-Feynman \cite{udrescu2020ai}, Feyn \cite{brolos2021approach, christensen2022identifying}, and AI-Descartes \cite{cornelio2023combining}—learn compact equations directly from data without assuming specific model structures. While effective in low-data regimes, they often struggle when systems involve differential or integral operators. To overcome these challenges, hybrid modeling approaches that integrate data-driven and physics-based priors have emerged. Notably, physics-informed neural networks (PINNs) \cite{raissi2019physics} and their variants (e.g., X-PINNs \cite{jagtap2021extended}) enforce known physical laws during training and have been combined with symbolic methods for gray-box system discovery \cite{kiyani2023framework}.
\textcolor{blue}{Here by gray-box modeling we  refer to hybrid frameworks that combine known mechanistic equations with neural components that learn unknown or uncertain dynamics from data; importantly, the missing part of the governing equations cannot be discovered directly from data, since the data describe the solution of the differential system rather than its underlying operators, and must therefore be inferred indirectly through physics-constrained learning.} Additional contributions include random projection neural networks \cite{de2018greedy, de2021interaction} and symbolic-layer least-squares models \cite{kokturk2021symbolic}, which improve scalability and interpretability.

Early gray-box modeling using neural networks \cite{rico1994continuous} has since evolved into frameworks for applications including phase-field models, optogenetics, and metabolic systems \cite{kemeth2023black, lovelett2019partial}. In biology, parameter inference from noisy and sparse data has motivated the development of systems biology-informed neural models \cite{yazdani2020systems, daneker2023systems}. AI-Aristotle \cite{Ahmadi2024} exemplifies this trend by integrating domain-decomposed PINNs with symbolic regression~\cite{raissi2024physics} to perform gray-box modeling under data constraints. Similarly, CMINNs \cite{daryakenari2025cminns} combine (fractional) PINNs with compartmental models to estimate time-varying PK/PD parameters, enabling the modeling of non-standard drug behaviors such as anomalous diffusion, delayed response, and resistance. These advances offer flexible representations for inverse problems in biomedical systems where parameter constancy assumptions may fall short. 

Despite these developments, training PINNs and their variants remains a key bottleneck. Their composite loss functions—balancing physics residuals and data misfit—result in non-convex optimization landscapes prone to local minima and slow convergence. This is further exacerbated in systems pharmacology, where datasets are sparse, unbalanced, and noisy, and the parameters of interest are often time-dependent or embedded within nonlinear dynamics. Consequently, optimization algorithms play a crucial role in the performance and reliability of these models.

Current training methods rely heavily on first-order optimizers such as stochastic gradient descent and Adam \cite{34,35}, which, while scalable, often plateau prematurely. High-order methods like Newton’s method offer faster convergence by leveraging curvature but are constrained by memory and computation costs due to the Hessian. To mitigate this, quasi-Newton methods such as BFGS and L-BFGS approximate the inverse Hessian using rank-one or rank-two updates \cite{41,42,43}, and have been adapted to PINN training with some success \cite{elham}. Recent studies have also explored adaptive loss weighting, warm-up strategies, exponential feature scaling, and sequential training as means to improve convergence and stability \cite{Ahmadi2024, URBAN2025113656}.

In this work, we systematically examine the optimization landscape of gray-box inverse problems in systems biology. Specifically, we investigate the impact of various optimizers—including adaptive first-order methods with learning rate schedulers and second-order methods like BFGS with backtracking and trust-region line search—on training dynamics and solution quality. Additionally, we assess the role of numerical precision (float32 vs. float64) in convergence behavior and final model accuracy. Our findings provide actionable insights for developing stable and interpretable physics-informed models in pharmacology and broader biological contexts.

The paper is organized as follows. In Section 2, we present the methodology, detailing the architectures of PINNs and the proposed tanh-cPIKANs, along with their respective formulations and training strategies. In Section 3,  we introduce representative pharmacokinetics and pharmacodynamics models used for evaluation, both of which represent gray-box inverse problems characterized by data sparsity, non-uniqueness, and ill-posedness. In Section 4, we provide a systematic investigation into representation and optimization choices for gray-box discovery. We compare the performance of MLP-based PINNs and KAN-based architectures, explore the influence of numerical precision (single vs. double) on convergence and stability, and evaluate the effectiveness of various optimizers and learning rate schedulers. 
Moreover, we assess the computational cost and accuracy trade-offs associated with using the JAX framework. Finally, in Section 5 we summarize the key findings, emphasizing the advantages of the proposed tanh-cPIKAN architecture to improve training stability and accuracy for gray-box system identification.
\section{Methodology}

Physics-Informed Networks (PINs) are  models that embed physical laws—typically represented by ordinary or partial differential equations (ODEs/PDEs)—directly into the training process of function approximators such as neural networks or Kolmogorov--Arnold Networks. This is accomplished through automatic differentiation, which enables the computation of derivatives required to enforce consistency with the governing equations. The term ``PINs'' broadly encompasses both Physics-Informed Neural Networks (PINNs) and Physics-Informed Kolmogorov--Arnold Networks (PIKANs). These models are particularly well-suited for solving inverse problems, where the objective is to infer unknown quantities---such as parameters, boundary conditions, or unobserved inputs---by leveraging observed data alongside prior physical knowledge.

Inverse problems are commonly encountered in various fields such as engineering, medical imaging, geophysics, and pharmacometrics, where the objective is to infer unobservable characteristics of a system from indirect measurements. Traditional techniques often rely on iterative optimization or complex data assimilation methods. However, PINs offer a more flexible and scalable approach by leveraging deep learning frameworks that are specifically designed to incorporate the underlying physics of the system.

In the context of PINs for gray-box discovery, the goal is to estimate the missing component of the equations---the unknown dynamics of the model ($f(t)$)---in a physical system described by a PDE or ODE. A typical formulation involves a network that aims to minimize a loss function, which is based on the physical model and the available data. Concretely, suppose we have a differential operator
\[
\mathcal{N}[\hat{u}(t; \theta), f(t)] = 0, 
\quad t \in [0, T],
\]
where $\mathcal{N}$ represents the known portion of the governing equations, $\hat{u}(t; \theta)$ is the predicted solution (dependent on the network parameters~$\theta$), and $f(t)$ is the unknown part of the model. 

To enforce these physics-based constraints, we typically select a set of collocation points $\{t_n\}_{n=1}^N$ in the domain $[0,T]$. At each collocation point, we substitute the neural network outputs (and their derivatives, obtained via automatic differentiation) into $\mathcal{N}$ to measure the physics mismatch, i.e., how closely the PDE/ODE residual is satisfied. Simultaneously, we incorporate a data mismatch term that compares the predicted solutions $\hat{u}(t;\theta)$ with observed data $u_{\text{data}}$ at measurement times $\{t_m\}_{m=1}^{M}$. Combining both terms yields a total loss function of the form: 
\[
\mathcal{L}(\theta) \;=\; 
\underbrace{\frac{1}{M}\sum_{m=1}^{M} \bigl\|\hat{u}(t_m;\theta) - u_{\text{data}}(t_m)\bigr\|^2}_{\text{data mismatch}}
\;+\;
\underbrace{\frac{1}{N}\sum_{n=1}^{N}\bigl\|\mathcal{N}[\hat{u}(t_n;\theta), f(t_n)]\bigr\|^2}_{\text{physics mismatch}}.
\]
In the inverse problem setting, the objective is to determine  $f(t)$ such that the predicted solutions $\hat{u}(t; \theta)$ match the observed data $u_{\text{data}}$ while remaining consistent with the governing equations. By automatically differentiating $\hat{u}$ with respect to $t$, PINs ensure that the ODE/PDE constraints are satisfied across the collocation points, merging both data-driven and physics-based insights in one framework.

\begin{figure}[!h]
    \centering
    \includegraphics[width=0.75\textwidth]{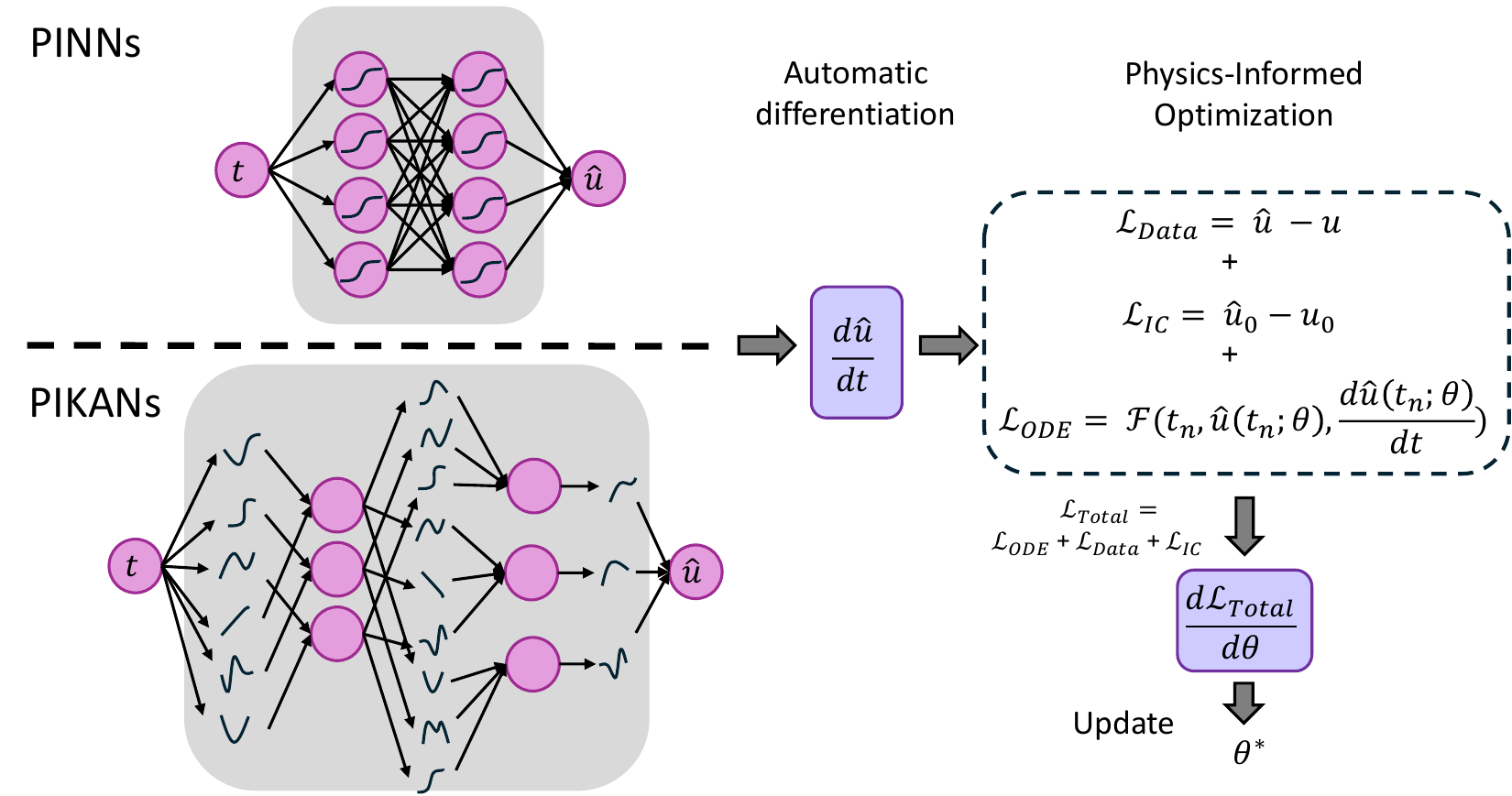}
    \caption{PINs: Physics-Informed Networks. Here $u$ and $u_0$ are observables at $t>0$ and $t=0$, respectively. $\hat{u}$ and $\hat{u}_0$ are inferred field from PINNs or PIKANs at t>0 and $t=0$, respectively. }
    \label{fig:network}
\end{figure}

In this study, we compare the performance of two different representation networks for $\hat{u}(t;\theta)$ and $f(t)$: the conventional multilayer perceptron (MLP) and the Kolmogorov--Arnold Network (KAN). In the following sections, we provide details on these two architectures and discuss their respective advantages for capturing the missing dynamics in gray-box discovery scenarios.
Schematics of the architectures of PINNs and PIKANs are shown in Fig. \ref{fig:network}.

\subsection{Kolmogorov-Arnold Networks Representation Model}
Motivated by the Kolmogorov–Arnold representation theorem, Kolmogorov–Arnold Networks (KANs) have been introduced as a structured alternative to conventional Multi-Layer Perceptrons (MLPs) \citep{liu2024kan}. This theorem states that any continuous function defined on a compact \( d \)-dimensional domain, \( f : [0,1]^d \rightarrow \mathbb{R} \), can be decomposed into a finite combination of nested univariate continuous functions:

\begin{equation}
f(x_1, \dots, x_d) = \sum_{q=0}^{2d} g_q\left( \sum_{p=1}^{d} \psi_{p,q}(x_p) \right),
\label{eq:kart1}
\end{equation}

where \( \psi_{p,q} \in C([0,1]) \) are inner functions acting on individual inputs, and \( g_q \in C(\mathbb{R}) \) are outer functions. This decomposition motivates the design of KANs, where each layer approximates such functional compositions in a data-driven, learnable manner.

Physics-Informed KANs (PIKANs) for forward modeling were systematically explored in \citep{shukla2024comprehensive, review1,ahmadi202new}, where each layer of the network is defined by:

\begin{equation} 
\bm{z}^{(l)} = \sum_{i=1}^{H}\Phi_i\left(\sum_{j=1}^{D}\phi_{i,j}(z^{(l-1)}_{j})\right), 
\label{eq:pikan}
\end{equation}

where \( \bm{z}^{(l-1)} = \{z_1^{(l-1)}, \cdots, z_H^{(l-1)}\} \) is the input vector to layer \( l \), \( \phi_{i,j} \) are trainable inner univariate functions, and \( \Phi_i \) are outer functions. Here, \( H \) denotes the number of nodes in layer \( l \), and \( D \) denotes the degree of the polynomial. The structure above is directly motivated by the theoretical form of Eq.~\ref{eq:kart1}.

Among various architectural variants, \citep{shukla2024comprehensive, ss2024chebyshev} introduced cPIKANs, which use \textit{Chebyshev polynomials} to parameterize the inner and outer functions. A typical inner function in this setting is given by:

\begin{equation}
\phi(\zeta; \theta) = \sum_{n=0}^{D} c_n T_n(\zeta),
\end{equation}

where \( \zeta \in [-1, 1] \) is the input, \( T_n \) is the \( n \)-th Chebyshev polynomial defined recursively as \( T_0(\zeta) = 1, T_1(\zeta) = \zeta, T_n(\zeta) = 2\zeta T_{n-1}(\zeta) - T_{n-2}(\zeta) \), and \( \theta = \{c_n\} \) are learnable coefficients. This spectral representation has been shown to improve robustness and computational efficiency. \textcolor{blue}{In the context of physics-informed training, our experiments demonstrate that Chebyshev polynomial bases outperform Jacobi variants: despite the improved expressivity reported in Ref.~\cite{AFZAL} for broader machine-learning tasks (rather than PINNs), our parameter-matched results in ~\ref{appendixnew} show that Jacobi-based KANs incur substantially longer runtime—due to their parameter-dependent recurrence coefficients, additional normalization factors, and larger computational graphs that hinder kernel fusion—while offering no accuracy improvement over the Chebyshev-based PIKANs.
}

\vspace{1em}
\noindent
\textbf{Our Modification.} In this study, we modified the KAN architecture to improve its stability, particularly for inverse problems. The modified version of cPIKAN, referred to as tanh-cPIKAN, is defined mathematically as follows.

\[
\sigma(x) = \tanh(x)
\]

\begin{equation}
f(\boldsymbol{x}) \approx 
W \cdot \sigma \left(\sum_{i_{L}=1}^{H_{L}} \sum_{n_L=1}^{D}  c_{n_L} T_{n_L} \left( \sigma \left(  
    \cdots  
      \sum_{i_1=1}^{H_1}\sum_{n_1=1}^{D} c_{n_1} T_{n_1} \left(\sigma \left( 
        \sigma \left( \sum_{i_0=1}^{H_0}\sum_{n_0=1}^{D} c_{n_0} T_{n_0}(\sigma(x))
      \right) 
    \right) 
    \right) 
  \right)
  \right) 
  \cdots
  \right) 
\end{equation}


In our architecture, \(D\) denotes the degree of the polynomials. The set \(\{ H_i \}_{i=0}^L\) represents the number of nodes in the \(i^{\text{th}}\) layer, where \(L\) is the total number of layers. The trainable parameters of our network, \(\theta\), include both the Chebyshev coefficients \(\{ c_j \}_{j=0}^L\) and the final weight matrix \(W\). The matrix \(W\) has dimensions equal to the number of nodes in the last hidden layer multiplied by the number of outputs.

In our network, applying a second \(\tanh\)---i.e., computing \(\tanh(\tanh(x))\)---further contracts the activation range from \([-1, 1]\) to approximately \([-0.76, 0.76]\) (since \(\tanh(1) \approx 0.7616\) and \(\tanh(-1) \approx -0.7616\)). 
%
This additional non-linear mapping, even on inputs already confined to \([-1, 1]\), refines feature representations and stabilizes learning by further squashing extreme values, which smooths the gradient flow (as gradients near the saturation limits decrease) and provides implicit regularization. This controlled contraction of the activation range ensures that the inputs to subsequent layers—such as those employing Chebyshev polynomial expansions—remain within an optimal domain, thereby enhancing numerical stability and improving the model's approximation accuracy. Additionally, by applying \(\tanh\) at the last hidden layer, all nodes output are mapped to \([-1, 1]\), and the scaling multiplication by \(W\) is then used to adjust these values to the required range for our problem.

\begin{figure}[!h]
    \centering

    \begin{subfigure}[t]{0.4\textwidth}
        \centering
        \includegraphics[width=\textwidth]{  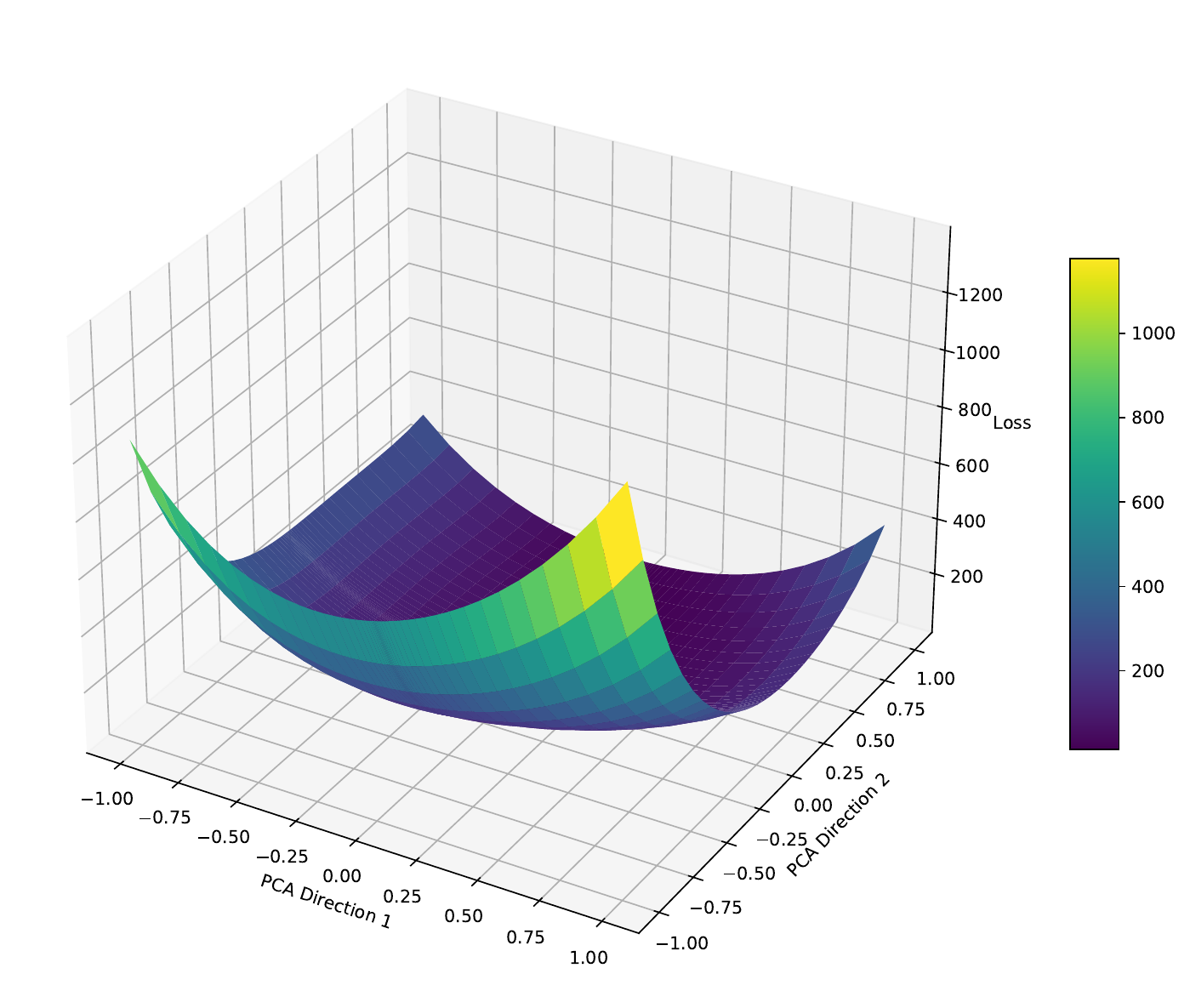}
        \caption{Loss landscape of cPIKANs}
    \end{subfigure}
    \hspace{0.02\textwidth}
    \begin{subfigure}[t]{0.4\textwidth}
        \centering
        \includegraphics[width=\textwidth]{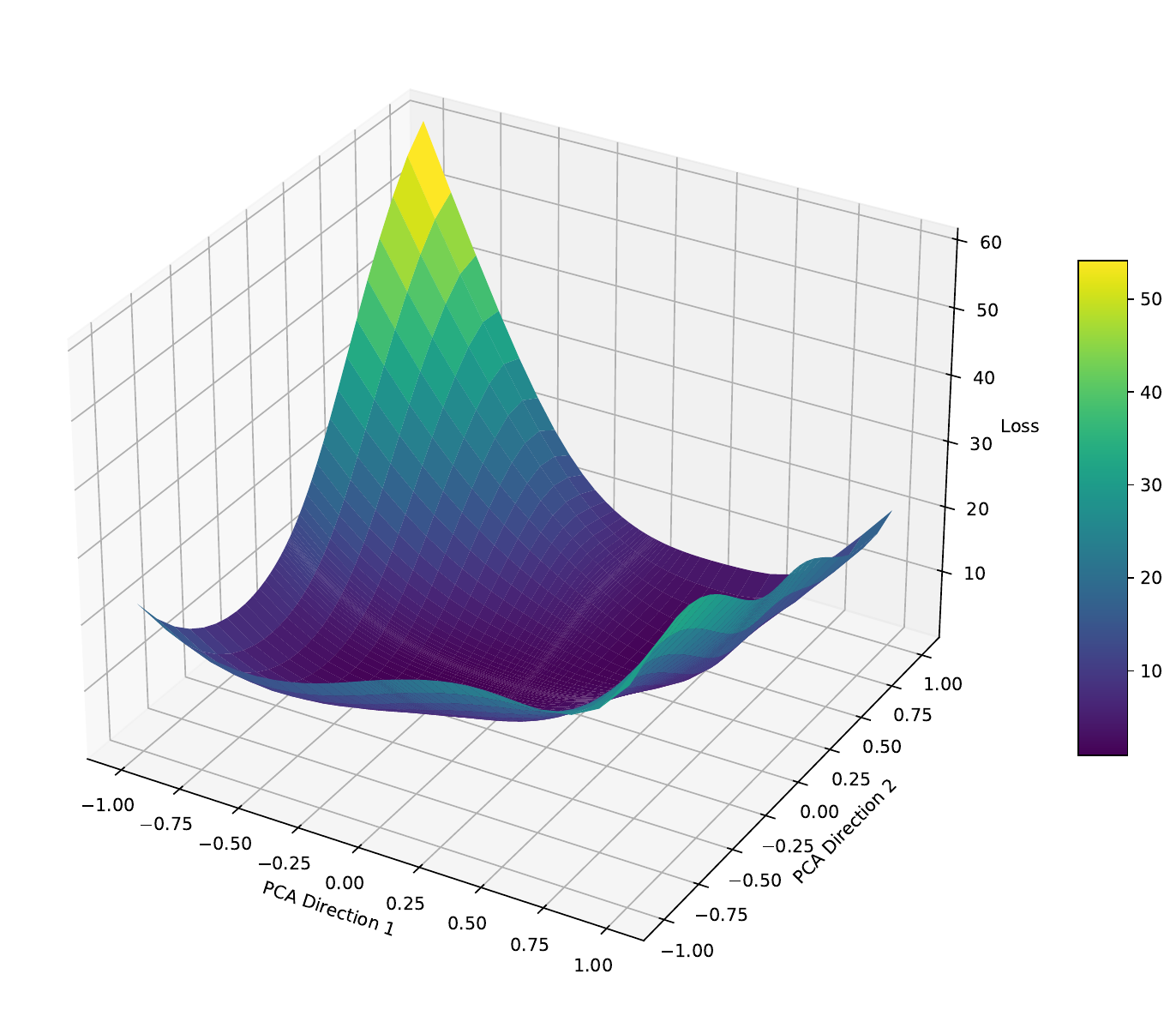}
        \caption{Loss landscape of tanh-cPIKANs}
    \end{subfigure}

    \caption{PK model: Comparison of loss landscapes between cPIKANs and tanh-cPIKANs in the PCA subspace. Both models are trained for the same number of iterations(70k), and parameter snapshots were collected every 100 epochs. Subfigures (a) and (b) illustrate the 3D loss surface reconstructed from the top two PCA directions of parameter evolution. Introducing outer nonlinearities in tanh-cPIKANs smoothens the loss surface and improves convergence and robustness.}
    \label{fig:loss_lnd}
\end{figure}

\textbf{Loss Landscape Analysis.}
To evaluate the optimization dynamics and convergence characteristics of our models, we conducted a PCA-based loss landscape analysis on the pharmacokinetics model. Parameter snapshots were collected every 100 epochs during training, and the top two principal components were extracted to define a low-dimensional subspace capturing the dominant directions of parameter evolution. We visualized the loss surface in this subspace for both the cPIKANs and the tanh-cPIKANs models, each trained for the same number of iterations, 
see Fig. \ref{fig:loss_lnd}. The loss landscape for the cPIKANs model displays a relatively steep and high-loss surface with values ranging approximately from 200 to over 1200. This indicates that the optimization is highly sensitive to perturbations along specific PCA directions, with sharp gradients suggesting that even small deviations in the parameter space lead to significant increases in the loss. In contrast, the tanh-cPIKANs model exhibits a much smoother and lower-loss landscape, with loss values spanning roughly from 10 to 60 and featuring a well-defined convex basin centered around the final parameter set. The smoother geometry implies that the tanh-cPIKANs is less sensitive to parameter perturbations, enabling faster convergence and more robust optimization. These results underscore the benefits of introducing nonlinearity in the outer functions of the KAN structure. By transforming the loss surface into a smoother, well-behaved landscape, tanh-cPIKANs facilitate more efficient and stable training trajectories compared to standard cPIKANs.

\subsection{Multi-Layer Perceptron Representation Model}
A Multi-Layer Perceptron (MLP) is a widely used neural network architecture composed of stacked layers, where each layer applies a linear transformation followed by a pointwise nonlinearity. Concretely, for layer $l$,
\begin{equation}
\mathbf{z}^{(l)} 
\;=\; 
\sigma\!\Bigl(W^{(l)} \,\mathbf{z}^{(l-1)} \;+\; \mathbf{b}^{(l)}\Bigr),
\end{equation}
where $\mathbf{z}^{(l-1)}$ is the input to the layer, $W^{(l)}$ and $\mathbf{b}^{(l)}$ are the trainable weight matrix and bias vector, respectively, and $\sigma(\cdot)$ is an elementwise activation function such as $\tanh$ or ReLU. Unlike Kolmogorov--Arnold Networks, which explicitly use a sum of univariate functions for each layer, MLPs aggregate inputs through matrix multiplications, thus mixing coordinates in a single step. Despite the structural differences, MLPs also serve as universal approximators, achieving considerable success across a range of machine learning tasks.

The results so far for the forward problems regarding which representation is better, MLP or KAN, especially in the context of PINs are mixed and, in fact, problem-dependent \cite{shukla2024comprehensive}.

\subsection{Types of Error in Physics-Informed Networks}

\begin{figure}[!h]
    \centering
    \includegraphics[width=0.75\textwidth]{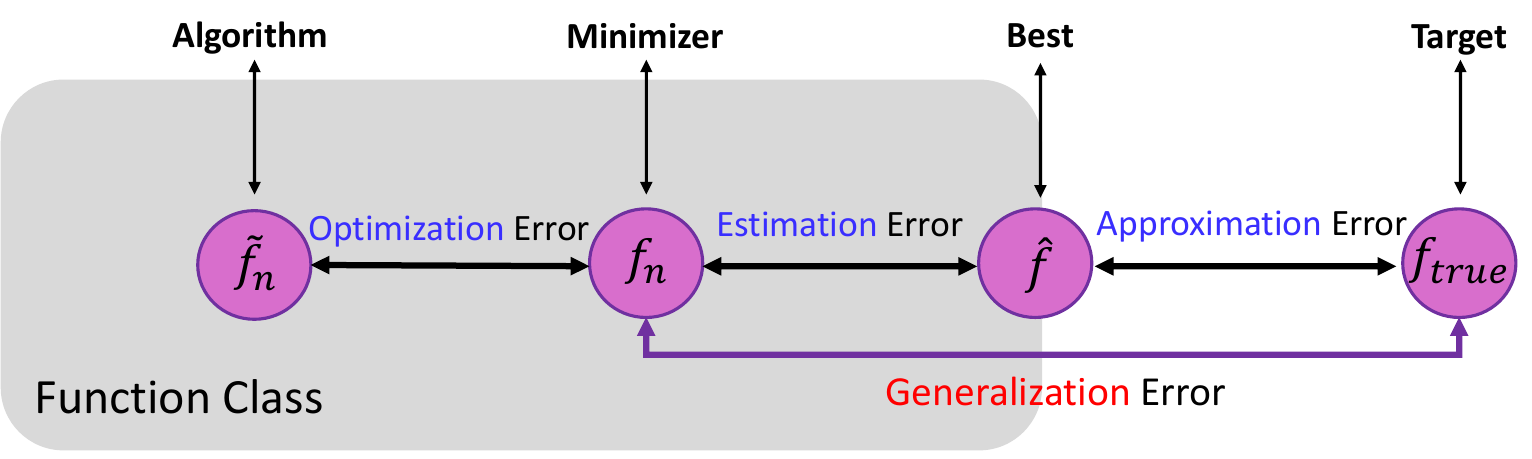}
    \caption{Types of error in Physics-Informed Networks (PINNs and PIKANs).}
    \label{fig:error}
\end{figure}

PINs exhibit various types of errors, which can lead to reduced accuracy compared to traditional numerical methods. Figure~\ref{fig:error} illustrates the sources of error during network training. The primary sources of these errors are as follows:

\subsubsection{Approximation (Representation) Error}
The first type of error in PINs is approximation (or representation) error, which for MLPs stems from the \textit{Universal Approximation Theorem}. This theorem asserts that a feedforward neural network with sufficient neurons and a nonlinear activation function can approximate any continuous function $ f: \mathbb{R}^n \to \mathbb{R}^m $ on a compact subset of $ \mathbb{R}^n $. However, practical limitations such as network architecture, training time, and computational resources can lead to approximation errors in solving real-world physics-informed problems.
KANs differ from MLPs by placing learnable activation functions on the edges (weights) instead of the nodes. This eliminates linear weight matrices and replaces each weight with a learnable 1D function, in this work modeled as a Chebyshev polynomial. The nodes in KANs simply sum incoming signals in each layer. Unlike MLPs, KANs provide an exact representation but the approximation error is due to the parameterization of the univariate functions.

In this study, we compare the KAN and MLP representation models  using two complementary strategies: (i) matching the number of hidden layers and the number of nodes/neurons per layer, and (ii) keeping the total number of trainable parameters approximately equal. In the first setup, this results in the KAN architecture having more parameters in the training loop, leading to a longer convergence time. In the second approach, enforcing an equal number of parameters requires the MLP to adopt a deeper architecture while maintaining the same number of neurons per layer as the KAN. Our goal is to determine which approach performs better in capturing complex dynamics.

\subsubsection{Estimation Error}
The second type of error is the estimation or generalization error, which originates from the finite sampling of collocation points used to calculate the physics loss during training iterations. Increasing the number of collocation points can reduce this error by enforcing the physical constraints with higher resolution; however, this improvement comes at the cost of increased computational complexity. Thus, a trade-off exists between accuracy and computational efficiency. In this work, we do not analyze this error, as we have hyper-tuned the parameters, including the number of collocation points, and kept it fixed for both the PINNs and tanh-cPIKANs.

\subsubsection{Optimization Error}

The optimization error arises during the training PINNs and PIKANs. Both methods involve solving non-convex optimization problems, which are inherently challenging because the models can become trapped in local minima. When this occurs, further training or additional data may not significantly improve the model's performance, as it reaches a saturation point. The efficiency of optimization algorithms is largely determined by their update rules, which are influenced by hyperparameters like the learning rate that control their behavior.

In this context, let the differentiable loss function be denoted by $ \ell: \mathbb{R}^d \to \mathbb{R} $, and its gradient as $ \nabla \ell(\bm{\theta}) $, where $ \bm{\theta} \in \mathbb{R}^d $ represents the model parameters. The goal of the optimization process is to find a parameter vector that minimizes the loss function, ideally reaching a local minimum.

\subsubsection*{First-Order Optimization Methods}
First-order optimization methods solve this problem by iteratively updating the parameter vector $ \bm{\theta}_k$ in a way that converges towards a local minimum. These methods rely on the evaluation of the loss function $ \ell $ and its gradient $ \nabla \ell $. The updates are guided by an update rule $ \mathcal{M} $, which determines the next iterate $ \bm{\theta}_{k+1} $ based on the current history of iterates, gradients, and function values:
\[
\mathcal{H}_k = \{(\bm{\theta}_s, \nabla \ell(\bm{\theta}_s), \ell(\bm{\theta}_s))\}_{s=0}^k.
\]
The next iterate is computed as:
\[
\bm{\theta}_{k+1} = \mathcal{M}(\mathcal{H}_k, \phi_k),
\]
where $ \phi_k $ is a set of hyperparameters, such as the learning rate, and the process starts from an initial value $ \theta_0 \in \mathbb{R}^d $.

\subsubsection*{Second-order (quasi-Newton) optimization method}
Unlike first-order approaches, which update the parameters of neural networks by using  the gradient of the loss function, second-order optimization techniques perform the update using curvature characteristics of the loss function by computing the Hessian, \( \mathbf{H}(\bm{\theta}) = \nabla^2 \ell(\bm{\theta)} \), which quantifies the local second-order behavior of the objective function. By leveraging both gradient and Hessian information, second-order methods can achieve improved convergence rates—particularly in scenarios where the loss surface exhibits poor regularity. A canonical form of second-order update reads as follows
\begin{align}\label{eq:opt_update}
\bm{\theta}_{k+1} = \bm{\theta}_k + \alpha_k \mathbf{p}_k,
\end{align}
where $\mathbf{p}_k$ is search direction pointing towards a region of lower function values on a high dimension loss function and expressed as
\begin{align}\label{eq:pk}
\mathbf{p}_k = -(\mathbf{H}_k)^{-1} \nabla \ell(\bm{\theta}_k).
\end{align}

To construct a second-order optimizer, we need a combination of algorithms to compute both \( \alpha_k \) and \( \mathbf{p}_k \). In \autoref{eq:pk}, the inverse of the Hessian matrix \( (\mathbf{H}_k)^{-1} \) is required for updating the parameters. However, computing the Hessian matrix can be computationally expensive for high-dimensional problems. Moreover, away from the solution, the Hessian may not be positive definite and can become ill-conditioned. To balance computational efficiency and convergence performance, quasi-Newton methods \cite{nocedal1999numerical} are employed. These methods iteratively approximate the Hessian matrix using only first-order derivative information, avoiding the need for explicit second-derivative computations. In this work, we use the BFGS update formula to compute \( \mathbf{H}_k \), which is expressed as \cite{AlBaali1993,AlBaali2005}
\begin{equation}\label{eq:Bkp1BFGS}
\mathbf{H}_{k+1}^{-1}=\frac{1}{\tau_k}\left[\mathbf{H}_k^{-1}-\frac{\mathbf{H}_k^{-1} \boldsymbol{y}_k \otimes \mathbf{H}^{-1}_k \boldsymbol{y}_k}{\boldsymbol{y}_k \cdot \mathbf{H}^{-1}_k \boldsymbol{y}_k}+\phi_k \boldsymbol{v}_k \otimes \boldsymbol{v}_k\right]+\frac{\boldsymbol{s}_k \otimes \boldsymbol{s}_k}{\boldsymbol{y}_k \cdot \boldsymbol{s}_k},
\end{equation}
where \( \mathbf{s}_k = \bm{\theta}_{k+1} - \bm{\theta}_k \) represents the change in the iterates, and \( \mathbf{y}_k = \nabla \ell({\bm{\theta}_{k+1}}) - \nabla \ell (\bm{\theta}_k) \) represents the corresponding change in gradients and  
\(\boldsymbol{v}_k=\sqrt{\boldsymbol{y}_k \cdot H_k \boldsymbol{y}_k}\left[\frac{\boldsymbol{s}_k}{\boldsymbol{y}_k \cdot \boldsymbol{s}_k}-\frac{H_k \boldsymbol{y}_k}{\boldsymbol{y}_k \cdot H_k \boldsymbol{y}_k}\right]\). $\otimes$ represents the outer product. $\tau_k$ and $\phi_k$ are 
are defined as the scaling and the updating parameters respectively. By choosing $\tau_k=\phi_k =1$, we recover the original BFGS algorithm \cite{nocedal1999numerical}. The choice of $\tau_k$ and $\phi_k$ improves the conditioning of $\mathbf{H}_k$ and in this study we also compare the results against the Self-Scaled BFGS algorithm \cite{AlBaali2005}, for which $\tau_k$ and $\phi_k$ are chosen as
\[
\tau_k=\min \left(1, \frac{-\boldsymbol{y}_k \cdot \boldsymbol{s}_k}{\alpha_k \boldsymbol{s}_k \cdot \nabla \ell \left(\boldsymbol{\theta}_k\right)}\right),~\phi_k=1.
\]

Once \( \mathbf{H}_k \) is computed, the next step is to compute \( \alpha_k \). The value of \( \alpha_k \) depends on the specific method being used and serves a similar role to the learning rate in first-order optimization methods, where only the gradient of the loss function is used to update the parameters. It is essential to choose \( \alpha_k \) carefully to ensure the accuracy of the local gradient estimation and promote reduction of the loss function, without hindering convergence by being too small. Consequently, the step size \( \alpha_k \) is typically chosen through inexact line search procedures, which maintain the positive definiteness of \( \mathbf{H}_k \) by imposing certain restrictions on \( \alpha_k \).

In this study, we examine two line search approaches: 1) Backtracking, and 2) Trust region. The backtracking line-search strategy is based on the Armijo condition \cite{armijo1966minimization}, which is formulated as follows,
\begin{align}\label{eq:armijo}
\ell\left(\bm{\theta}_k + \alpha_k \textbf{p}_k \right) \leq \ell \left(\bm{\theta}_k \right) + c_1 \alpha_k \textbf{p}_k^{\top}  \nabla \ell(\bm{\theta}_k).
\end{align}

In the backtracking strategy, the step length is initially set to a reasonable value ( $\bar{\alpha}$; not too small). If the condition is satisfied, the step size \( \alpha_k \) is accepted (i.e., \( \alpha_k = \bar{\alpha} \)), and the algorithm proceeds to the next iteration. If the condition is not met, the step size is reduced by multiplying it by a factor \( \rho < 1 \), and the process is repeated until the Armijo condition is satisfied. Additionally, the following condition is imposed along with \eqref{eq:armijo},
\begin{align}\label{eq:btrack}
\textbf{p}_k^{\top} \ell(\bm{\theta}_k) \le 0.
\end{align}

The second line search method we investigated is based on the trust-region approach. In a trust-region algorithm the direction and the step-length are calculated at the same time. In the trust trust-region approach, we define a region, known as the trust region, around the current point \( \bm{\theta}_k \) and look for a point within this region that reduces $\ell_k$. To accomplish this, a trust-region algorithm solves the following subproblem:
\begin{align}
\begin{aligned}
&\min _{\mathbf{p}} \nabla \ell_k^{\top} \mathbf{p}+\frac{1}{2} \mathbf{p}^{\top} \mathbf{H}_k \mathbf{p}\\
&\text { subject to: }\\
&\|\bm{p}\| \leq \Delta_k
\end{aligned}
\end{align}
where $\Delta_k$ is the radius of the trust-region, and $\textbf{H}_k$ is some approximation of the Hessian matrix at $\bm{\theta}_k$. Then, the proposed step \( \textbf{p}_k \) is evaluated. If it results in a significant improvement in the objective function, the step is accepted and the trust-region can be expanded. If the step yields poor results, it is rejected, and the trust-region is reduced. 
Finally, the iterate $\bm{\theta}_k$ is updated simply as 
\[
\bm{\theta}_{k+1} = \bm{\theta}_k + \textbf{p}_k,
\]
Therefore, combining $\textbf{H}_k$ and $\alpha_k$  as described by \autoref{eq:opt_update} results into a quasi-Newton methods which achieve superlinear convergence by progressively improving the Hessian approximation \( \mathbf{H}_k \).


\subsubsection*{Optimization Error Analysis in this Study}
In this study, we compare the performance of PINNs and PIKANs under different optimization strategies. We assess optimization error in four distinct settings:

\begin{enumerate}
    \item \textbf{First-Order Optimizers:} We evaluate the performance of several first-order optimizers applied to both PINNs and tanh-cPIKANs, aiming to identify optimal configurations for gray-box system identification tasks. Our analysis considers the combined effects of optimizer choice, initial learning rate, and learning rate schedulers. 
    
    \item \textbf{Second-Order Optimizers:} We evaluate the effectiveness of using a second-order optimizer throughout the entire training process, investigating its impact on convergence speed and accuracy for both PINNs and tanh-cPIKANs.
    
    \item \textbf{Hybrid Optimization Strategy:} We also explore a hybrid training approach in which the model is initially trained with a first-order optimizer and then switched to a second-order optimizer. This strategy aims to leverage the fast initial convergence of first-order methods and the fine-tuning capabilities of second-order optimization.

    \item \textbf{Precision Settings}: We investigate the impact of single-precision versus double-precision settings during training. Previous computational experiments \cite{double} have shown that single-precision can achieve performance comparable to double-precision under certain conditions, particularly when the line search algorithm effectively identifies improvements. In this study, we aim to explore the circumstances under which double-precision settings outperform single-precision in solving inverse problems.
\end{enumerate}


\subsubsection*{Robustness to Observational Noise}
To complement our optimization analysis, we evaluate the robustness of PINNs and tanh-cPIKANs when trained with noisy data. In real-world pharmacological settings, measurement noise is unavoidable; thus, it is critical to assess model stability under such perturbations.
\section{Problem Setup}
In this section, we introduce the mathematical models utilized to capture complex dynamics and uncover underlying system equations by employing different optimization techniques and PIN architectures. The first model represents a pharmacokinetics model, while the second describes a nonlinear pharmacodynamic system designed to simulate response to multi-dose chemotherapy treatment schedules. The primary objective is to address an inverse problem that is inherently ill-posed and potentially non-unique, especially under conditions of sparse data. By identifying missing components within these models, we enable a robust evaluation of various optimization strategies for gray-box discovery challenges in quantitative systems pharmacology. Furthermore, we conduct a comparative analysis of the performance of two distinct physics-informed frameworks: PINNs and tanh-cPIKANs.


\subsection{Pharmacokinetics Model}

One of the models utilized in this work is a single-dose compartmental pharmacokinetics model, which is described by the following system of ordinary differential equations:
\begin{equation}\label{eq:PK}
    \begin{aligned}
        \dfrac{dB}{dt} &= k_g G - k_b B, \\
        \dfrac{dG}{dt} &= -k_g G, \\
        \dfrac{dU}{dt} &= k_b B.
    \end{aligned}
\end{equation}

This model captures the time-dependent concentration of a drug across three compartments over a 50-hour interval. Initially, the drug is introduced into the gastrointestinal (GI) tract (compartment $G$), where it dissolves and enters the bloodstream (compartment $B$). The drug is subsequently eliminated through metabolic and excretory processes involving the liver, kidneys, and urinary tract (compartment $U$). 
The parameters $k_g = 0.72 \, h^{-1}$ and $k_b = 0.15 \, h^{-1}$ represent the rates at which the drug transitions from the GI tract to the bloodstream and is cleared from the bloodstream, respectively. For this study, the administered drug is modeled as $0.1 \, \mu g$ of tetracycline, an antibiotic.

In the gray-box approach, missing terms in the model are approximated based on available data. For the PK model, the unknown term appears in the right-hand side of the first ODE, which we represent as an unknown function $f(t)$:

\begin{equation}\label{eq:PK_f}
    \begin{aligned}
        \dfrac{dB}{dt} &= f(t),\\
        \dfrac{dG}{dt} &= -k_g G,\\
        \dfrac{dU}{dt} &= k_b B.
    \end{aligned} 
\end{equation}

\begin{figure}[!h]
    \centering
    \includegraphics[width=0.8\textwidth]{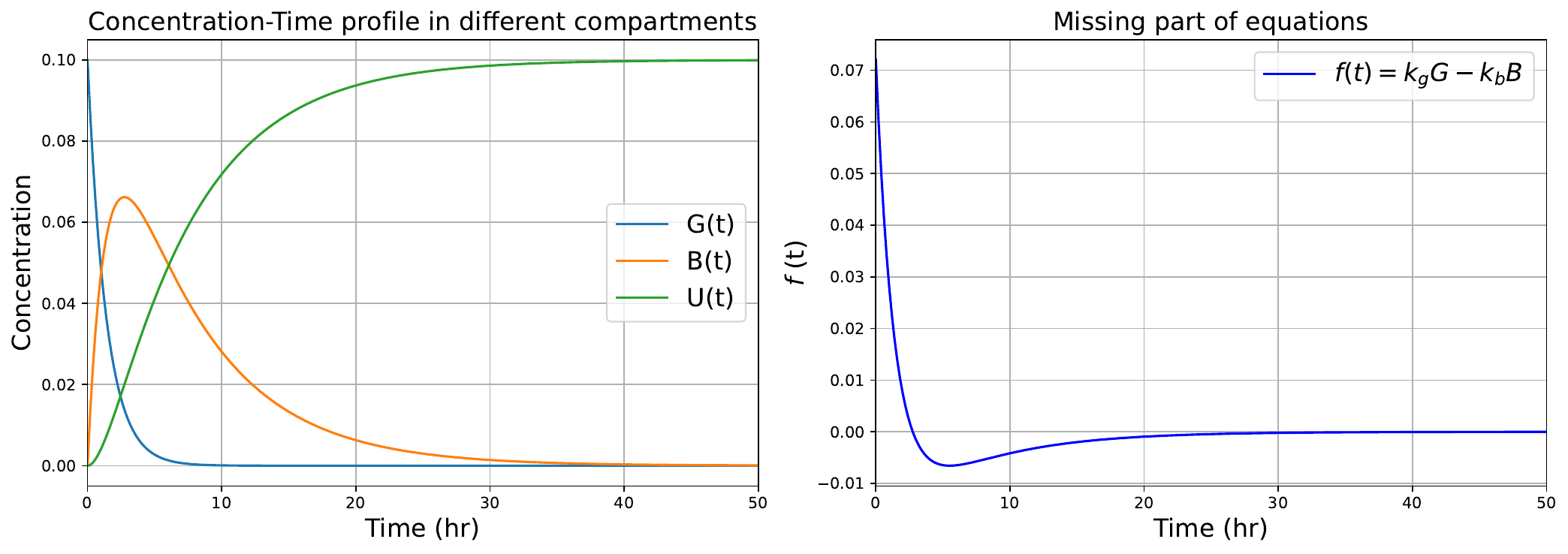}
    \caption{Pharmacokinetics model.}
    \label{fig:pk_model}
\end{figure}

The function $f(t)$ is determined by leveraging available data for $B$, $G$, and $U$, enabling the reconstruction of missing dynamics in the model. To generate synthetic data, we employed the \texttt{odeint} function from the \texttt{scipy.integrate} library, which leverages the LSODA algorithm. LSODA automatically switches between solvers depending on the stiffness of the system: it uses the \textit{Adams–Bashforth–Moulton} method (a multi-step predictor-corrector scheme) for non-stiff problems and the \textit{Backward Differentiation Formula (BDF)}, an implicit method, for stiff problems. The time span for simulation was set to 50 hours, with data sampled at 1-hour intervals.  Fig.~\ref{fig:pk_model} shows the solution of our pharmacokinetics ODE model, along with the missing component that we aim to discover using PINs.

\subsection{Pharmacodynamics Model}

The second model is a pharmacodynamics model used to describe the effect of a chemotherapy drug on cancer cell counts \cite{chemodata}. The time evolution of the cell count is governed by the following differential equation:
\begin{equation}\label{eq:PK_f}
    \frac{dN(t)}{dt} = (k_p - k_d(t,D)) N(t) \left(1 - \frac{N(t)}{\theta(D)}\right), 
\end{equation}
where $ N(t) $ denotes the cell count at time $ t $, $ k_p $ is the constant growth rate of the cells under untreated conditions, $ k_d(t, D) $ represents the rate of cell death, which is influenced by both the dosage $ D $ and the elapsed time $ t $, and $ \theta(D) $ is the carrying capacity dependent on the drug dosage, representing the upper limit of cell population that can be supported under a given dose.
The death rate $ k_d(t, D) $ can be further modeled using distinct functions, each having different dependencies on the dosage. 
\begin{equation}\label{eq:PK_f}
    \begin{aligned}
        k_d(t, D) &= k_{d,A}(D),\\
        k_d(t, D) &= k_{d,B}(D) r(D) t e^{1 - r(D) t}.
    \end{aligned} 
\end{equation}

Here, $ r(D) $ is a dosage-dependent parameter that affects the time decay of the drug’s influence. These parameters are cell-line specific. Previous studies have fitted the functions $ k_{d,A}(D) $ and $ r(D) $ separately using nonlinear least squares methods, and the final model prediction for each dosage is a weighted combination of the results from both functions \cite{chemodata}.
However, to simplify the computational model and avoid issues with identifiability, we assume that $ k_p $ remains constant and that fitting both $ k_d(D, t) $ and $ \theta(D) $ simultaneously would lead to an unidentifiable solution. Therefore, we reduce the model to the simpler form as follows:
\begin{equation}\label{eq:Pd}
    \frac{dN(t)}{dt} = F_D(N, t) N(t)
\end{equation}
In this case, $ F_D(N, t) $ is an unknown function that depends on time, which we refer to as the ``chemotherapy efficacy function'' in a multi-dose regimen.  While it could also incorporate both the number of cells $ N $ and time $ t $, we restrict it to depend solely on time for the purpose of data generation. Hence, $ F_D(N, t) $ is the function that we aim to discover in the inverse problem. To generate synthetic data for the chemotherapy pharmacodynamics model, we employed the same numerical solver used in the previous example—\texttt{odeint} from the \texttt{scipy.integrate} library—which utilizes the LSODA algorithm to handle both stiff and non-stiff systems. The simulation was performed over a time span of 600 hours post-treatment, with data sampled every 5 hours.

For the simulations, we set the carrying capacity to \( \theta(D) = 1 \), the intrinsic cell proliferation rate to \( k_p = 0.0345 \), the maximum drug-induced death rate to \( k_{d,B} = 0.03 \), and the drug efficacy decay rate to \( r(D) = 0.007 \). These parameter values were chosen to qualitatively reflect typical tumor response dynamics under chemotherapeutic treatment, consistent with prior modeling studies \cite{chemodata}. The time-varying, nonlinear, and dose-dependent efficacy function \( F_D(t) \) was designed to capture the dynamic impact of drug administration on cancer cell populations and serves as the ground truth to be identified through the inverse problem framework. 

\begin{figure}[!h]
    \centering
    \includegraphics[width=0.8\textwidth]{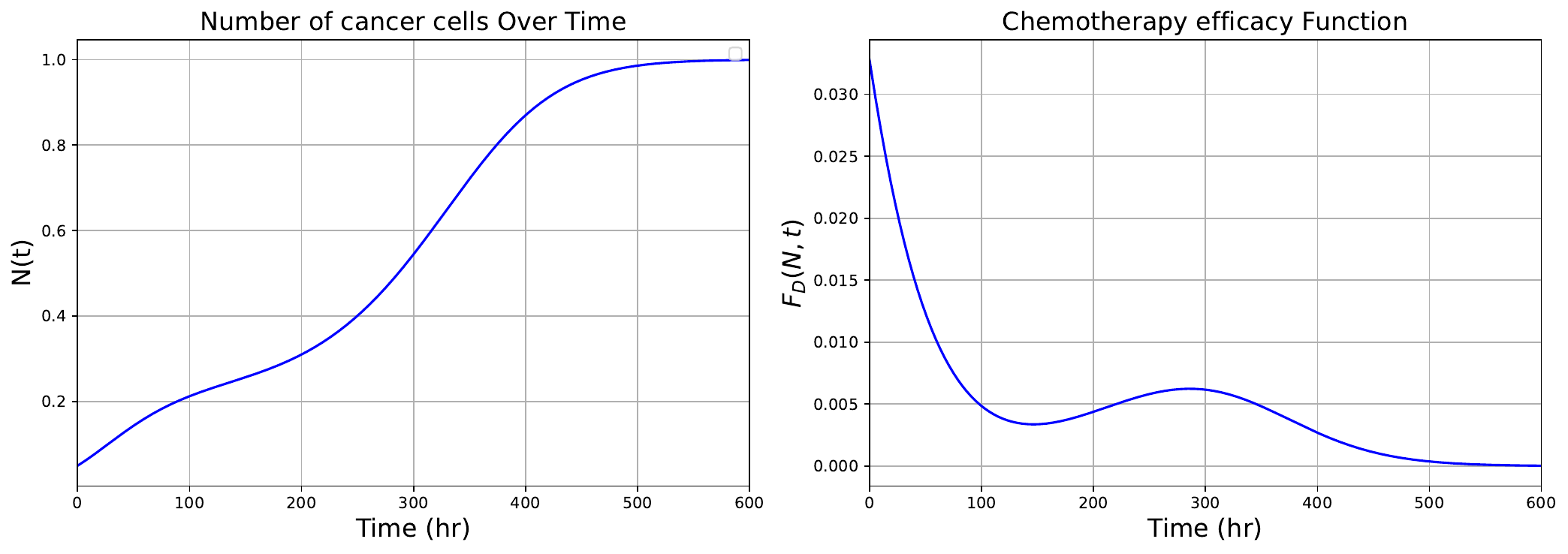}
    \caption{Pharmacodynamics model.}
    \label{fig:pd_model}
\end{figure}

\section{Results and Discussion}
In this section, we present and analyze the results of our comprehensive optimization study on physics-informed networks. The first subsection focuses on the pharmacokinetics (PK) model, evaluating the performance of various optimizers, representations, and training configurations across multiple case studies. The second subsection presents results for the pharmacodynamics (PD) model, specifically a chemotherapy drug-response system. We compare the performance of PINNs and PIKANs in gray-box system identification tasks. To generate training data, we solve the forward problems and sample the resulting trajectories to simulate sparse observations. Our analysis highlights the impact of architecture, optimizer type, and precision settings on learning efficiency and model accuracy under ill-posed conditions.
\subsection{Pharmacokinetics Model}
Choosing an appropriate optimizer and learning rate is one of the most crucial aspects of successfully training neural networks. These factors directly influence the stability, convergence speed, and final accuracy of the model. Despite their importance, selecting the appropriate combination of optimizer and learning rate remains a challenging task—especially in the context of PINNs and emerging variants such as PIKANs.

The learning rate, or step size, determines the magnitude of updates applied to model weights during backpropagation. A well-chosen learning rate can lead to fast and stable convergence, whereas poor choices may result in oscillations, divergence, or excessively slow training. Larger learning rates allow for faster learning in the early stages but risk overshooting minima or destabilizing training. Smaller learning rates promote finer convergence but can cause the optimizer to stagnate or get trapped in local minima. To mitigate these issues, we use learning rate schedulers, such as cosine decay, which start with a larger learning rate and gradually reduce it to refine the model during later training stages.

Optimizers play an equally important role. First-order methods like Adam, RAdam, and SGD with momentum differ in how they handle gradient updates, adaptivity, and stability. Choosing an unsuitable optimizer can significantly degrade model performance. In this work, all experiments were implemented in JAX, and we employed the Optax and Optimistix \cite{optimistix} libraries for their efficient, modular, and JAX-compatible optimization algorithms. 

For the pharmacokinetics model, both tanh-cPIKANs and PINNs used 2 hidden layers with 50 nodes/neurons per layer; this configuration resulted in approximately 10,300 trainable parameters for tanh-cPIKANs and 2,752 parameters for PINNs. \textcolor{blue}{In addition, in \ref{appendixnew} we provide results using architectures with a comparable number of parameters for both models. These experiments further demonstrate that the observed performance differences originate primarily from the representational and optimization characteristics of the tanh-cPIKAN architecture—namely, its smoother loss landscape and adaptive spectral representation—rather than differences in parameter capacity. By contrast, increasing the parameter count in PINNs (e.g., 50 neurons, 5 hidden layers, \(\sim 10{,}400\) parameters) did not yield proportional gains in accuracy.}

\subsubsection{Comparison of Model Performance Across First-Order Optimizers and Learning Rates}

\begin{figure}[!h]
    \centering
    \includegraphics[width=0.6\textwidth]{  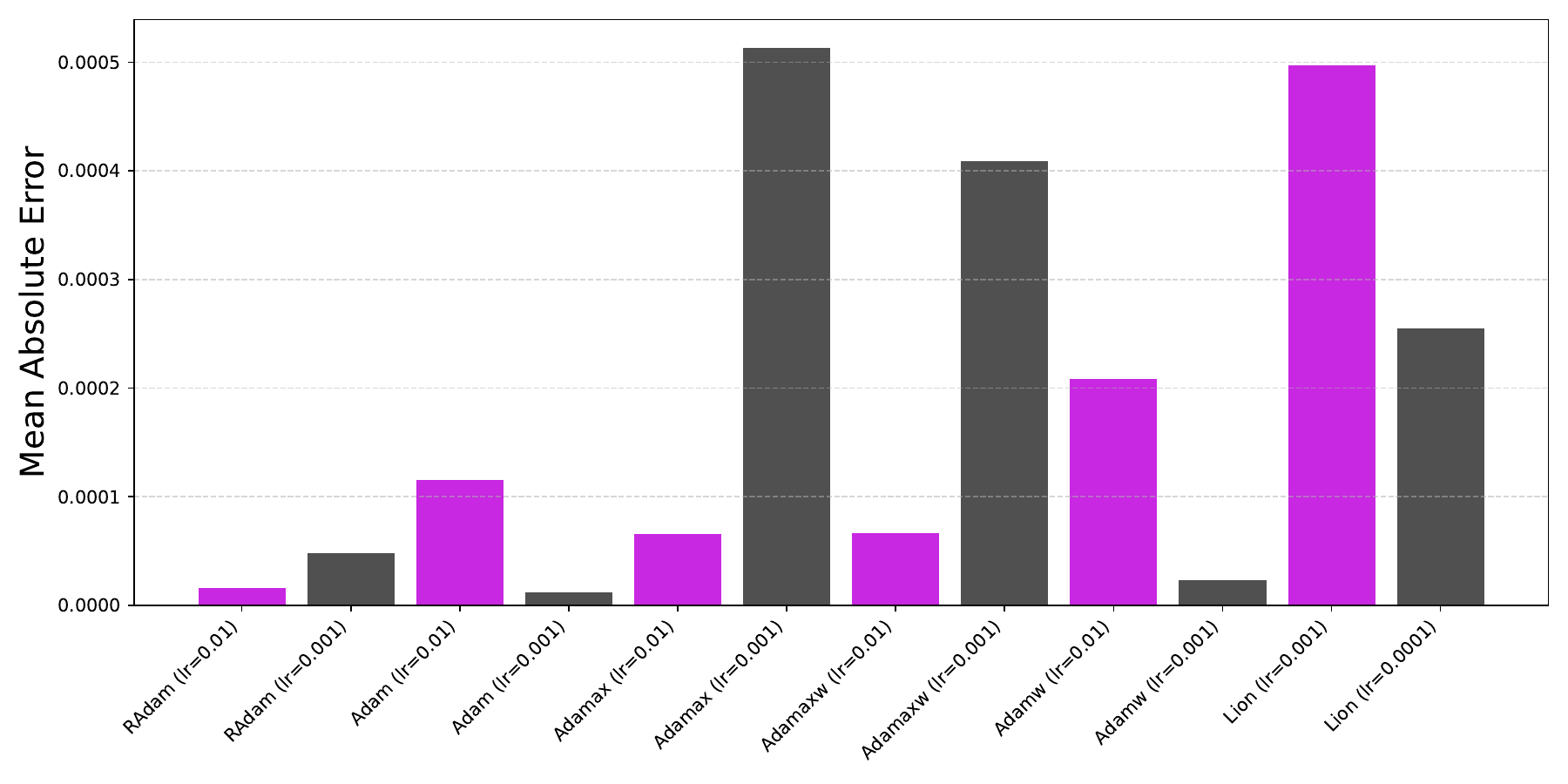}
    \caption{Mean Absolute Error of tanh-cPIKANs for discovering missing 
    dynamics across optimizers and initial learning rates.}
    \label{fig:pikan_opt}
\end{figure}

\begin{figure}[!h]
    \centering
    \includegraphics[width=0.6\textwidth]{  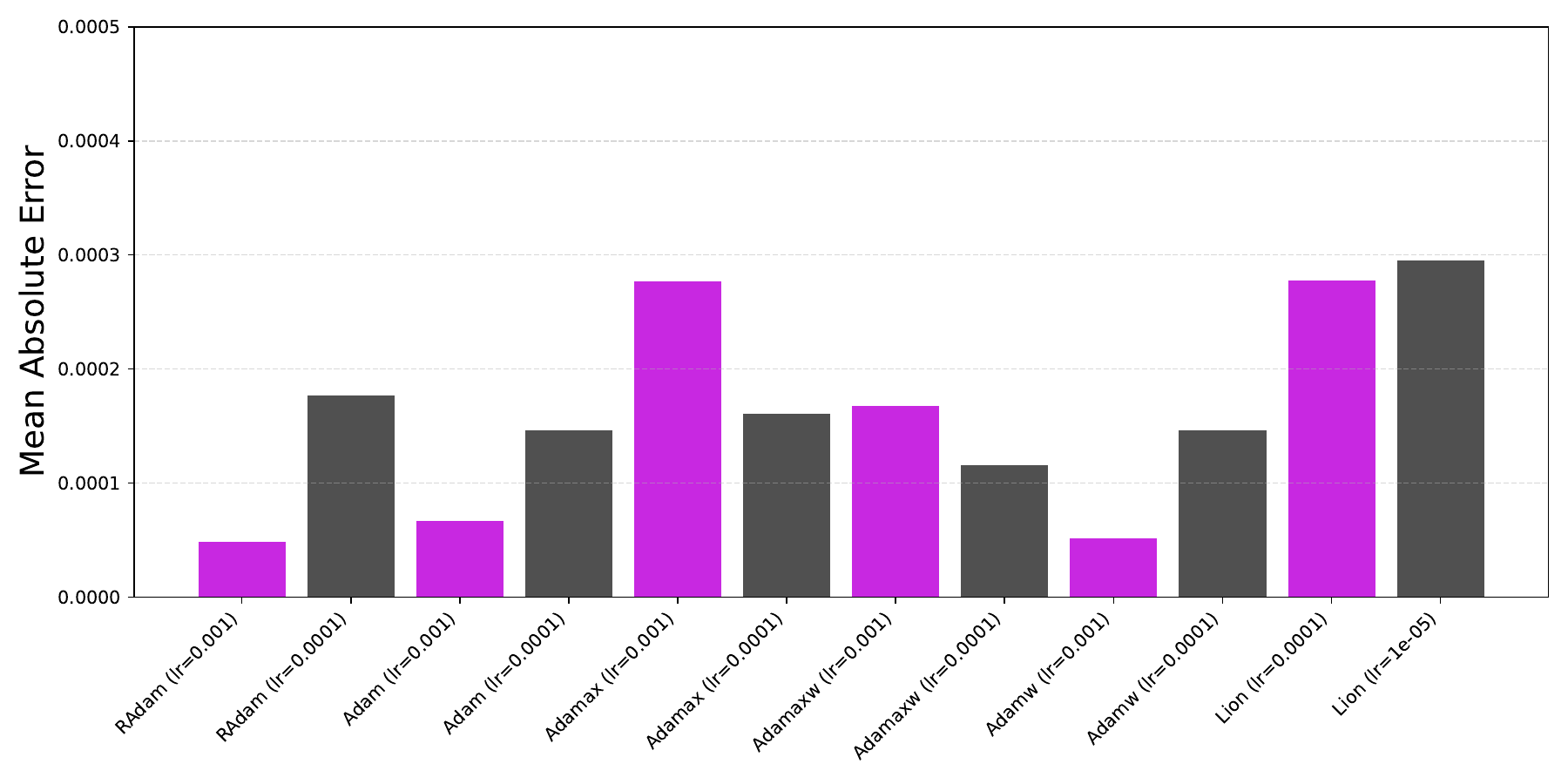}
    \caption{Mean Absolute Error of PINNs for discovering missing dynamics across optimizers and initial learning rates.}
    \label{fig:pinn_opt}
\end{figure}

This section explores the impact of first-order optimizers and initial learning rate choices on model performance, particularly in conjunction with a cosine learning rate scheduler. We evaluate both PINNs and tanh-cPIKANs on a pharmacokinetics model to assess their ability to identify the missing dynamic component $f(t)$ in Eq.~\ref{eq:PK_f}. Despite the use of a scheduler, the choice of initial learning rate significantly affects convergence speed and final accuracy. Figures~\ref{fig:pikan_opt} and~\ref{fig:pinn_opt} present the mean absolute error (MAE) of the discovered $f(t)$ compared to the numerical solution, highlighting the influence of optimizer and learning rate configurations in gray-box system identification.

Figure~\ref{fig:pikan_opt} presents the performance of tanh-cPIKANs across a range of first-order optimizers and initial learning rates. Among the tested configurations, RAdam with a learning rate of 0.01 and Adam with 0.001 achieve the lowest MAEs, indicating effective convergence and accurate recovery of the missing system component. In contrast, Lion yields substantially higher errors, particularly at learning rates of 0.001 and 0.0001, and performs even worse at 0.01, reflecting inefficient training dynamics. The performance of the Adamax optimizer also varies significantly with the choice of learning rate, further highlighting sensitivity to hyperparameter selection. These results underscore the strong dependence of tanh-cPIKAN performance on the optimizer–learning rate combination, even under a cosine scheduler, and emphasize the importance of careful tuning in gray-box system identification tasks.

The observed differences in optimizer performance can be attributed to their underlying update mechanisms and how they interact with the nonlinear architecture of cPIKANs. Adam and RAdam are first-order optimizers that incorporate second-moment estimates of the gradients—i.e., they track the variance of past gradients to adaptively scale learning rates per parameter. While this makes them more stable and adaptive in complex loss landscapes, it does not equate to true second-order optimization, which relies on curvature information from the Hessian. In contrast, Lion is a simpler optimizer that omits second-moment estimation and relies only on momentum and the sign of the gradient. Although more lightweight, Lion is more sensitive to learning rate settings and less robust in handling the sharp curvature and nonlinearity introduced by the Chebyshev-based representations in cPIKANs. As a result, Lion often requires much smaller learning rates to achieve competitive performance, as reflected in our experiments.

In contrast, figure~\ref{fig:pinn_opt} presents the results for PINNs, which were trained using learning rates one order of magnitude lower than those used for PIKANs (i.e., 0.001 and 0.0001). Overall, PINNs achieve lower MAEs and exhibit more consistent performance across different optimizers. RAdam and Adam continue to perform reliably, yielding stable results across both learning rates. Lion, similar to tanh-cPIKAN setting, required one order smaller learning rates (1e-4 and 1e-5) to reach comparable performance. The greater robustness of PINNs to optimizer and learning rate selection can be attributed to their simpler and more uniform architecture. Unlike tanh-cPIKANs, which employ adaptive Chebyshev polynomial-based transformations that introduce additional nonlinearity and parameter interactions, standard PINNs have a more stable gradient landscape and fewer trainable components, making optimization more predictable. This architectural simplicity likely contributes to their enhanced stability under varying training configurations in gray-box system identification tasks.

\begin{figure}[!h]
    \centering
    \includegraphics[width=0.6\textwidth]{  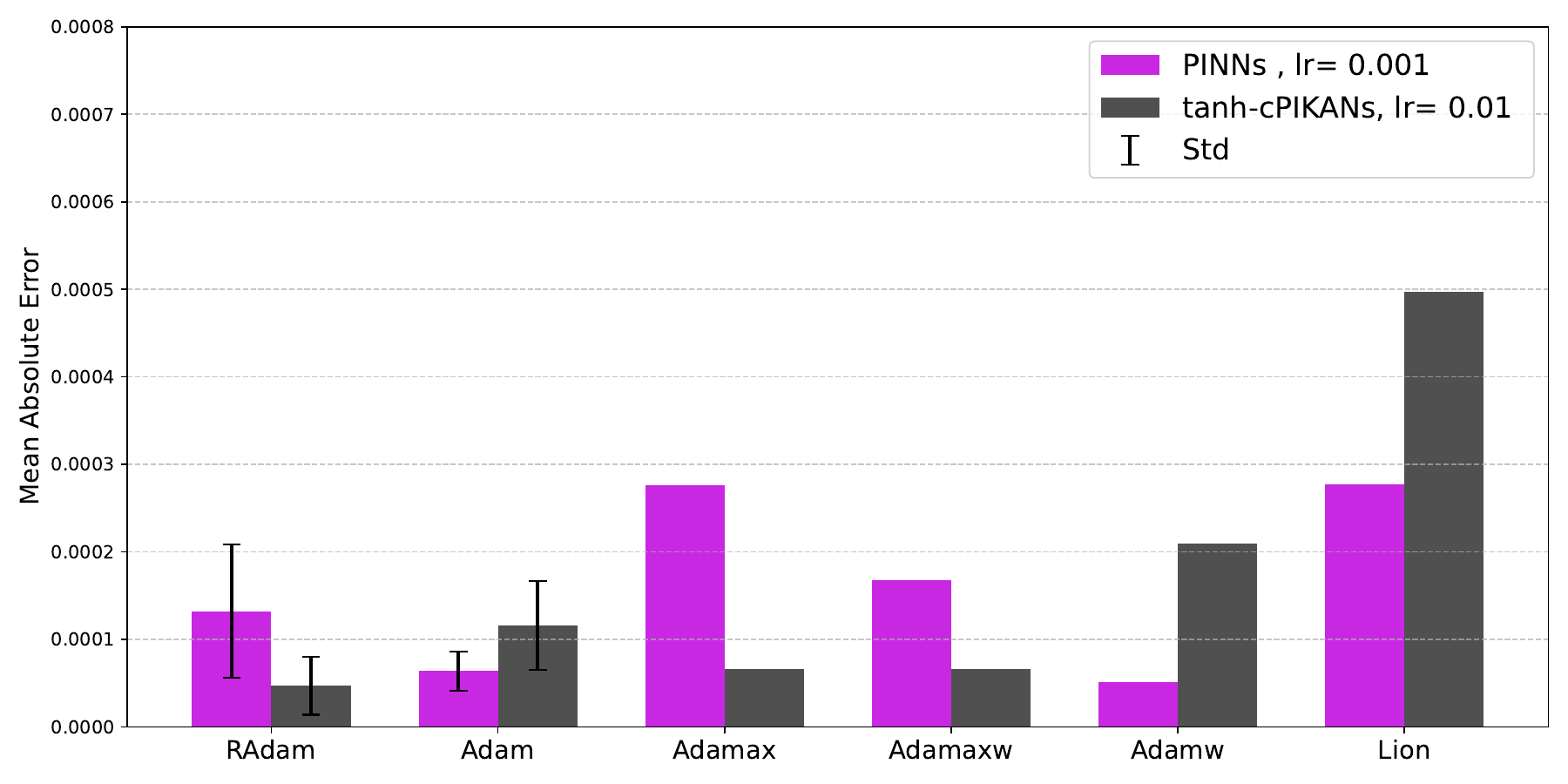}
    \caption{Comparison of PINNs and tanh-cPIKANs performance using different optimizers with a cosine scheduler. For the Lion optimizer, we used a learning rate one order of magnitude lower than the others: 0.0001 for PINNs and 0.001 for tanh-cPIKANs.}

    \label{fig:cos_pk}
\end{figure}

Given the differing nature of the two architectures, each exhibiting optimal performance at different learning rates, we tailored the training setup accordingly. For tanh-cPIKANs, we selected an initial learning rate of 0.01, as it consistently yielded superior results across most optimizers. In contrast, PINNs performed best with a smaller learning rate of 0.001, likely due to their simpler architecture and more stable training dynamics. To ensure a robust comparison, we conducted 10 independent trials for both the Adam and RAdam optimizers, which demonstrated strong and consistent performance across both models. Additionally, we trained the PINNs for 50,000 iterations and the PIKANs for 70,000 iterations. This difference reflects the higher parameter count and greater representational complexity of the tanh-cPIKAN architecture, which generally requires longer training to reach  and comparable results. The number of collocation points for the physics residuals is set to 500. The summarized results of this evaluation are presented in Figure~\ref{fig:cos_pk}.


The three best-performing optimizers for each model are as follows:

\begin{itemize} \item \textbf{tanh-cPIKANs:} RAdam, AdaMax, and AdaMaxW \item \textbf{PINNs:} AdamW, Adam, and RAdam \end{itemize}

These findings highlight RAdam as a consistently effective optimizer across both architectures, offering a strong balance between adaptivity and stability. Its rectified update mechanism appears particularly beneficial in navigating the more complex optimization landscape of tanh-cPIKANs, while still performing robustly in the smoother training regime of PINNs.


\subsubsection{Comparison of Different Schedulers for RAdam Optimizer}

\begin{figure}[!h] 
    \centering
    \includegraphics[width=0.6\textwidth]{  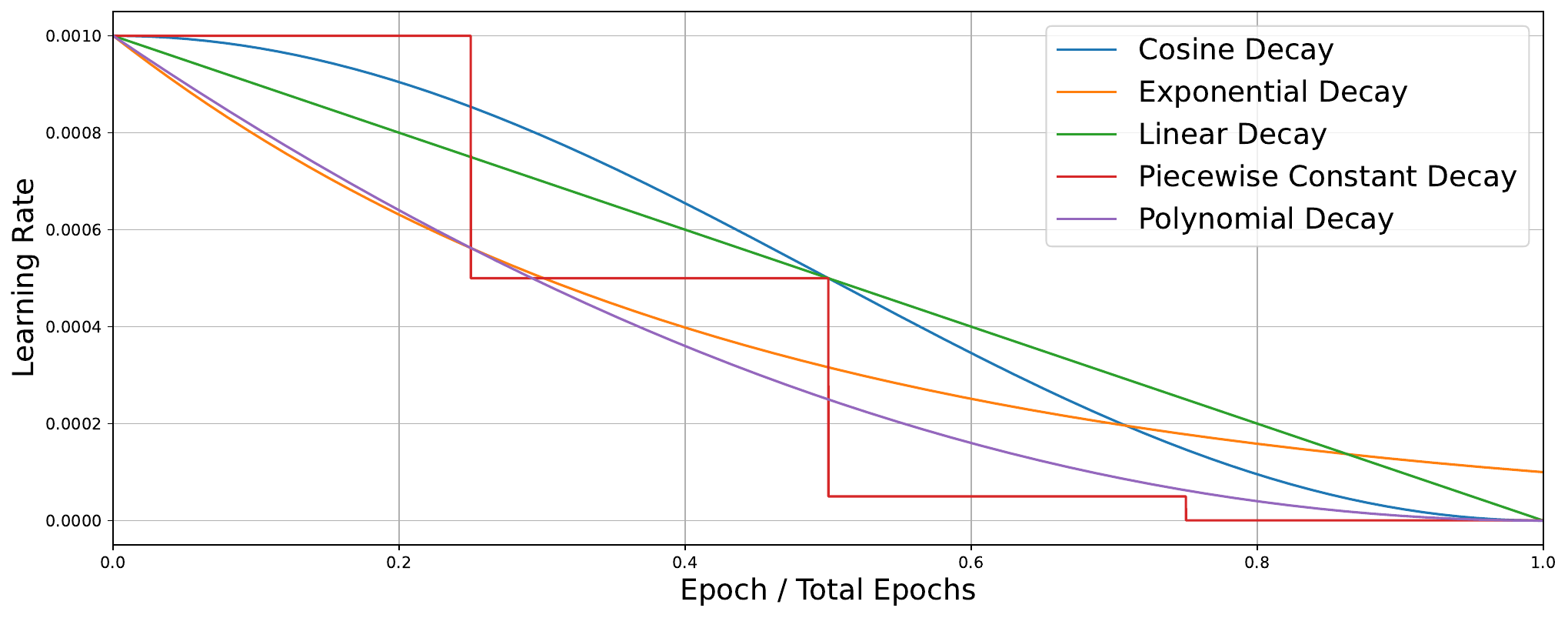} 
    \caption{Learning rate trajectories over the training process for different scheduling strategies (polynomial, exponential, cosine, linear, and piecewise), all starting from an initial learning rate of 0.001. The schedules were adjusted to ensure comparable learning rate ranges for a fair performance comparison between PINNs and PIKANs.}
    \label{fig:sch_lr} 
\end{figure}

\begin{figure}[!ht] 
    \centering
    \includegraphics[width=0.6\textwidth]{  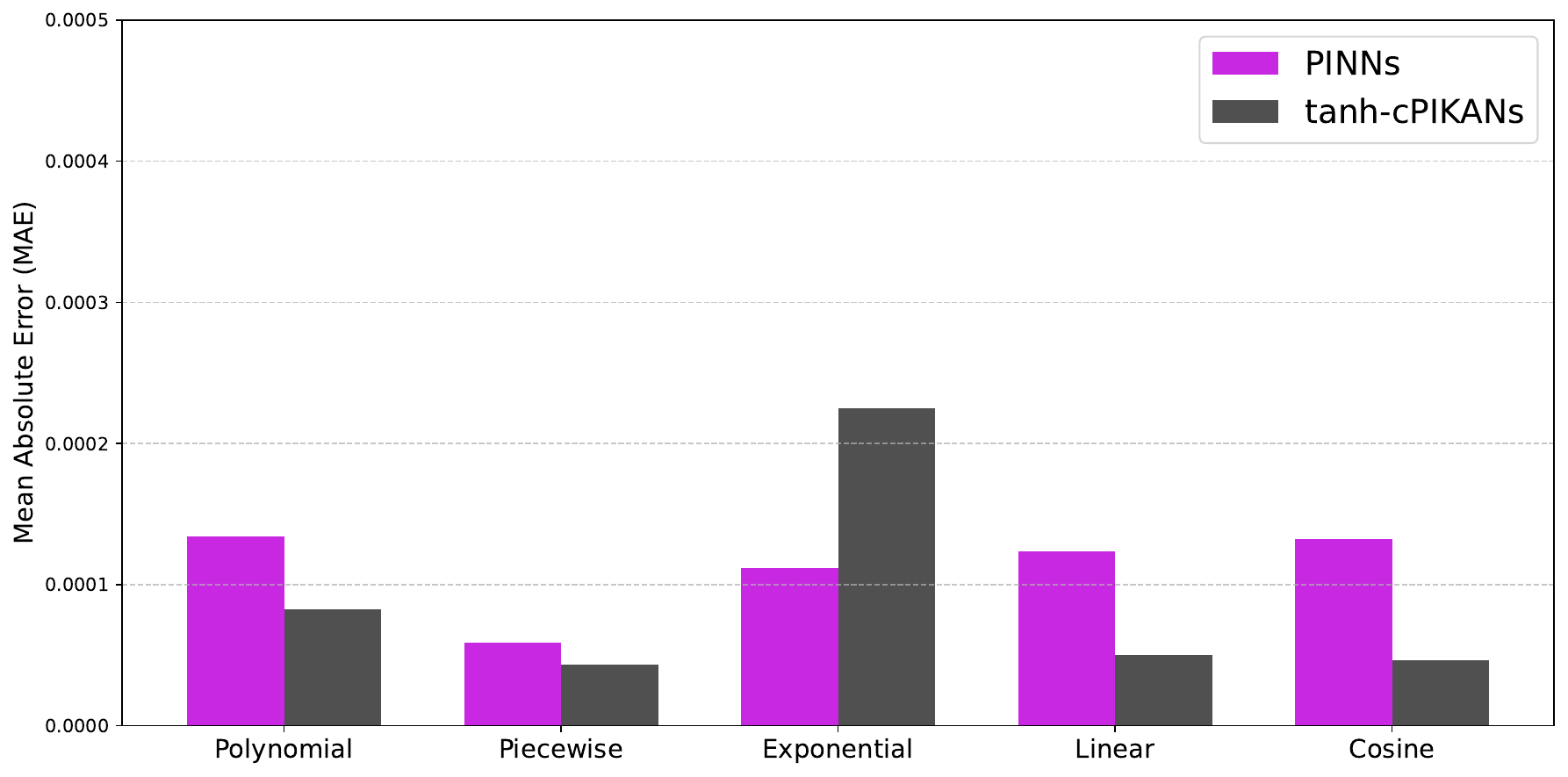} 
    \caption{Comparison of PINNs and tanh-cPIKANs accuracy in discovering the missing part of the pharmacokinetic model using different learning rate schedulers. All models were trained with the RAdam optimizer and an initial learning rate of 0.001, and scheduler hyperparameters were adjusted to ensure comparable learning rate ranges across representations.}
    \label{fig:sch_compare} 
\end{figure}

We evaluated the performance of PINNs and tanh-cPIKANs under various initial learning rates (0.01, 0.001, and 0.0001). The results indicated that tanh-cPIKANs achieved optimal performance with an initial learning rate of 0.01, whereas PINNs performed best with a starting rate of 0.001. To obtain comparable results when using the RAdam optimizer across different learning rate schedulers, we standardized the initial learning rate to 0.001 for both architectures, as summarized in Table~\ref{tab:cos_pk}. We also adjusted the scheduler-specific hyperparameters to keep the learning rate trajectories within a similar range during training. PINNs were trained for 50,000 iterations and tanh-cPIKANs for 70,000 iterations to account for their increased number of parameters and slower convergence. We examined several scheduling strategies, including polynomial decay, exponential decay, cosine annealing, linear decay, and piecewise constant decay. The learning rate profiles for each scheduler are shown in Figure~\ref{fig:sch_lr}, and their impact on the accuracy of recovering the missing component in the pharmacokinetics model is presented in Figure~\ref{fig:sch_compare}.

\begin{table}[!h]
\centering
\begin{tabular}{|c|c|c|c|}
\hline
\textbf{Model} & \textbf{Optimizer(\#itr.)} & \textbf{Error (MAE)} & \textbf{Comp. Time (s)} \\
\hline
tanh-cPIKANs & Adam(70k) & 5.25e-05 &307.56 \\
tanh-cPIKANs &  RAdam(70k) & 8.42e-05  & 303.23 \\
tanh-cPIKANs &  RAdam(50k)  & 1.30e-04 & 217.45 \\
\hline
PINNs & Adam(50) &  9.15e-05  & 108.89  \\
PINNs & RAdam(50) & 7.45e-05 & 94.875 \\
PINNs  & RAdam(70k) & 5.28e-05& 114.69 \\
\hline
\end{tabular}
\caption{PK model: Comparison of PINNs and tanh-cPIKANs using two first-order optimizers—Adam and RAdam. All experiments were conducted in \textit{single} precision with a cosine learning rate scheduler, starting from an initial learning rate of 0.001. The models were trained with varying numbers of iterations (\#itr.), as indicated in the table.}
\label{tab:cos_pk}
\end{table}

Figure~\ref{fig:sch_compare} compares the performance of various learning rate schedulers—polynomial, exponential, cosine, linear, and piecewise—applied to both PINNs and tanh-cPIKANs using the RAdam optimizer. Among these, the piecewise scheduler consistently outperforms the others for both architectures, achieving the lowest mean absolute error in recovering the missing pharmacokinetic component. This superior performance is likely due to its greater flexibility: the piecewise scheduler allows for manual control over when and how many times the learning rate changes (e.g., with 2, 3, or more scheduled steps). However, this flexibility comes at the cost of increased complexity in hyperparameter tuning. Determining the number, timing, and magnitude of learning rate drops is highly problem-dependent and can be difficult to optimize effectively without prior knowledge or extensive trial and error. In contrast, other schedulers like cosine and exponential offer more automated decay behavior with fewer hyperparameters to tune, making them easier to apply but slightly less performant in this setting. Linear and polynomial schedulers fall somewhere in between, showing moderate results with reasonable stability. Overall, while the piecewise scheduler yields the best accuracy, its practical application may require more manual effort and domain-specific tuning compared to smoother, more automated alternatives.

\subsubsection{Performance of PINNs and tanh-cPIKANs with Single Precision}

\begin{figure}[!h]
    \centering
    \includegraphics[width=0.8\textwidth]{  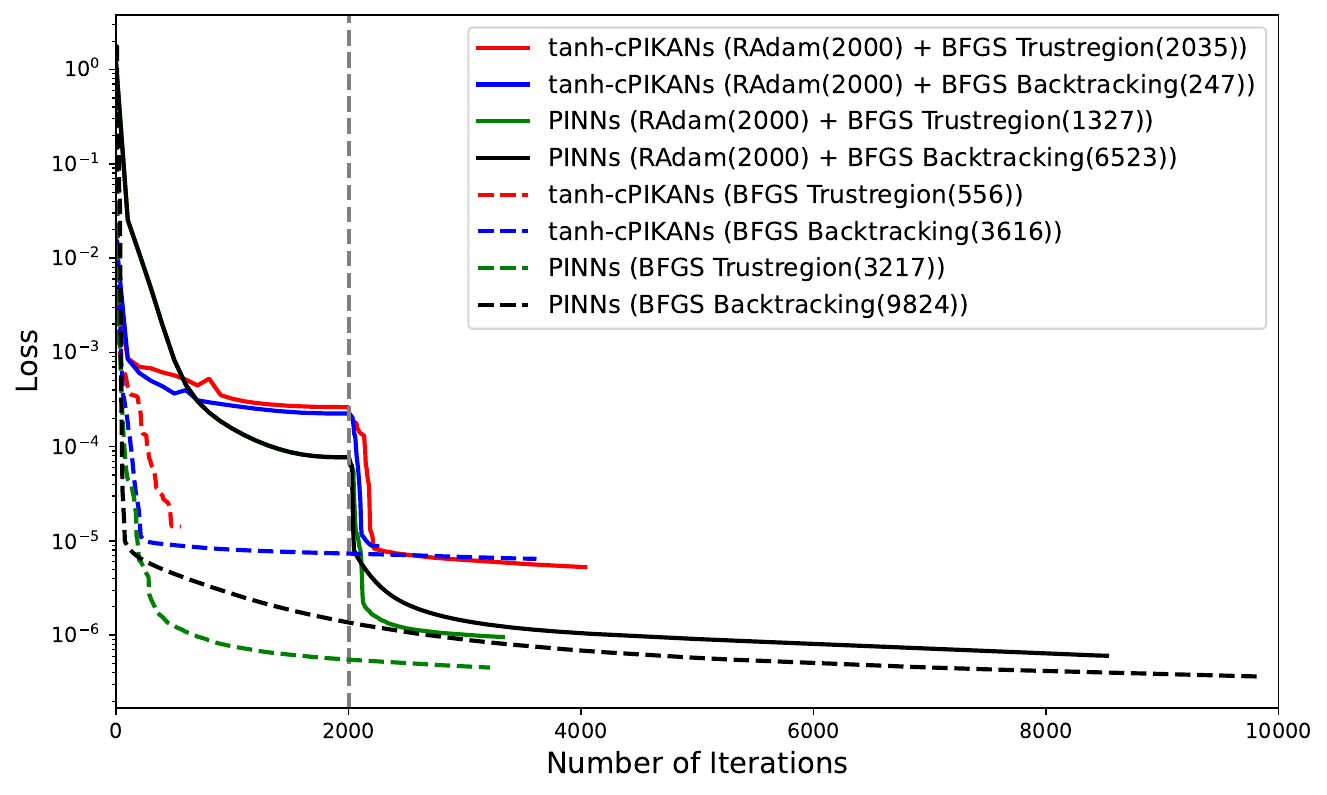} 
    \caption{PK model: Comparison of PINNs and tanh-cPIKANs performance with different optimizers for single precision.}
    \label{fig:single-8} 
\end{figure}

To assess the stability and efficiency of PINNs and tanh-cPIKANs under reduced numerical precision, we evaluated both architectures using single-precision (float32) arithmetic. As shown in Figure~\ref{fig:single-8} and Table~\ref{tab:PINNs_vs_PIKANs_single_precision}, both models successfully identified the missing dynamics, but with notable differences in accuracy and computational time across optimizer configurations.

In the course of our experimental investigations, we observed that first-order optimizers, such as RAdam, exhibited relatively consistent performance across both single- and double-precision settings. Their reliance on gradient-based updates and adaptive moment estimates renders them less sensitive to minor numerical fluctuations introduced by reduced precision. In contrast, second-order and hybrid optimizers demonstrated significantly improved performance under double precision. This enhancement is attributed to their reliance on more precise curvature information, such as Hessian approximations and gradient norms, during line search or trust-region updates. Second-order methods typically terminate when the norm of the gradient falls below a given threshold; thus, the level of numerical precision directly impacts their stopping criteria and convergence quality. In low-precision settings, these fine changes may be obscured, potentially causing premature convergence or unstable updates. These findings motivated our focused evaluation of hybrid and pure second-order strategies (e.g., BFGS with backtracking line search or trustregion line search methods) in single-precision mode.

When using hybrid optimization strategies—such as combining RAdam with BFGS—it is usually critical to include a warm-up phase with the first-order optimizer. This phase helps the network escape poor initializations and navigate early regions of the loss landscape where second-order curvature information is unreliable or uninformative. The timing of the transition from first- to second-order optimization is equally important: switching too early may lead the second-order optimizer to converge prematurely to suboptimal local minima, while switching too late may trap it in a narrow flat region where meaningful progress stalls. To explore this, we conducted an ablation study and found that training with RAdam for 2000 iterations before transitioning to a second-order optimizer (either \texttt{BFGS\_bck} or \texttt{BFGS\_trust}) provided the best balance for both PINNs and tanh-cPIKANs. This warm-up period allows the model to reach a more favorable region of the parameter space, enabling the second-order phase to refine the solution more effectively and achieve improved final accuracy.

 PINNs consistently achieved lower MAEs across all optimizer configurations, with the best result of $4.26 \times 10^{-4}$ obtained using a hybrid RAdam + \texttt{BFGS\_bck} strategy. In comparison, the best-performing tanh-cPIKAN configuration achieved an MAE of $7.44 \times 10^{-4}$ using \texttt{BFGS\_bck}, with significantly fewer iterations and reduced computational time.

Overall, PIKANs demonstrated shorter training times—particularly when using \texttt{BFGS\_trust}, which completed in just 7.07 seconds—while PINNs required longer runtimes due to their larger iteration counts, especially when combined with second-order optimizers. These results highlight a fundamental trade-off: while tanh-cPIKANs offer faster convergence with moderate accuracy, PINNs provide more stable and precise performance under reduced precision, particularly when leveraging well-tuned second-order or hybrid optimization strategies.

\begin{table}[!h]
\centering
\begin{tabular}{|c|c|c|c|}
\hline
\textbf{Model} & \textbf{Optimizer(\#itr.)} & \textbf{Error (MAE)} & \textbf{Comp. Time} \\
\hline
tanh-cPIKANs & BFGS\_bck(3.616K) & 7.44e-04 & 9.70 \\
tanh-cPIKANs & BFGS\_trust (0.556k) & 1.36e-03 & 7.07 \\
tanh-cPIKANs & RAdam(2k) + BFGS\_bck(0.247k) & 1.77e-03 & 19.06 \\
tanh-cPIKANs & RAdam(2k) + BFGS\_trust(2.035k) & 1.29e-03 & 23.81 \\
\hline
PINNs & BFGS\_bck(9.824k) & 4.57e-04 & 31.25 \\
PINNs & BFGS\_trust (3.217) & 4.81e-04 & 14.02 \\
PINNs & RAdam(2k) + BFGS\_bck(6.523K) & 4.26e-04 & 32.87 \\
PINNs & RAdam(2k) + BFGS\_trust(1.327k) & 5.18e-04 & 16.36 \\
\hline
\end{tabular}
\caption{PK model:Comparison of PINNs and PIKANs using different optimizers (hybrid or second-order) in terms of MAE and computational time. All experiments were performed in single precision, using a cosine learning rate scheduler with an initial learning rate of 0.001. “BFGS-bck” refers to the BFGS optimizer with a backtracking line search method, while “BFGS-trust” refers to the BFGS optimizer using a trust-region line search method.}
\label{tab:PINNs_vs_PIKANs_single_precision}
\end{table}

\subsubsection{Performance of PINNs and tanh-cPIKANs with Double Precision}

\begin{figure}[!h] 
    \centering
    \includegraphics[width=0.8\textwidth]{  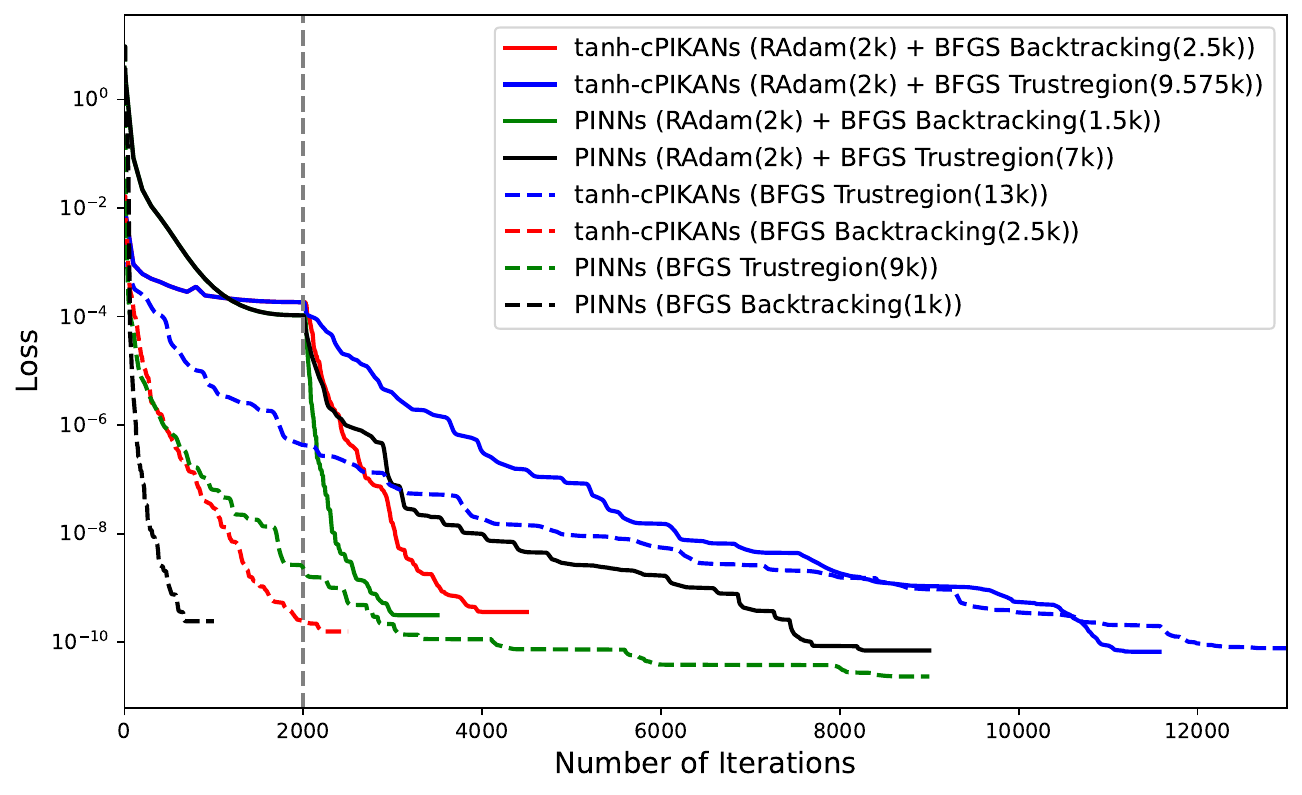} 
    \caption{PK model: Comparison of PINNs and PIKANs performance with different optimizers for double precision.}
    \label{fig:double_9} 
\end{figure}

To evaluate the benefits of higher numerical accuracy, we assessed the performance of PINNs and tanh-cPIKANs using double-precision (float64) arithmetic. The results, summarized in Figure~\ref{fig:double_9} and Table~\ref{tab:PINNs_vs_PIKANs}, show that both architectures benefit significantly from the increased precision, particularly when paired with second-order or hybrid optimization strategies.

\begin{table}[!h]
\centering
\begin{tabular}{|c|c|c|c|}
\hline
\textbf{Model} & \textbf{Optimizer} & \textbf{Error (MAE)} & \textbf{Comp. Time} \\
\hline
tanh-cPIKANs & BFGS\_bck(2.5K) & \textbf{5.92e-06} & \textbf{43.19} \\
tanh-cPIKANs & BFGS\_trust (13k) & 6.23e-06 & 244.14 \\
tanh-cPIKANs & RAdam(2k) + BFGS\_bck(2.5k) & 1.01e-05 & 51.44 \\
tanh-cPIKANs & RAdam(2k) + BFGS\_trust(9.576k) & 7.17e-06 & 179.22 \\
\hline
PINNs & BFGS\_bck(1K) & \textbf{2.26e-05} & \textbf{28.23} \\
PINNs & BFGS\_trust (9k) & 3.06e-05 & 218.53 \\
PINNs & RAdam(2k) + BFGS\_bck(1.5K) & 6.45e-05 & 51.50 \\
PINNs & RAdam(2k) + BFGS\_trust(7k) & 3.93e-05 & 187.29 \\
\hline
\end{tabular}
\caption{PK model: Comparison of PINNs and tanh-cPIKANs using different optimizers (hybrid or second-order) in terms of MAE and computational time. All experiments were performed in double precision, using a cosine learning rate scheduler for RAdam optimizer with an initial learning rate of 0.001. “BFGS-bck” refers to the BFGS optimizer with a backtracking line search method, while “BFGS-trust” refers to the BFGS optimizer using a trust-region line search method.}
\label{tab:PINNs_vs_PIKANs}
\end{table}

Tanh-cPIKANs demonstrated the best overall performance, achieving a minimum MAE of $5.92 \times 10^{-6}$ using \texttt{BFGS\_bck}, and outperforming all other configurations in both accuracy and efficiency. Notably, this setup required only 2.5k iterations and 43.19 seconds of training time. Other configurations, such as \texttt{BFGS\_trust} or hybrid RAdam + BFGS variants, also produced competitive results with slightly increased error and runtime. In contrast, while PINNs showed improved accuracy over their single-precision counterparts, they did not match the accuracy levels reached by tanh-cPIKANs. The best-performing PINN configuration achieved an MAE of $2.26 \times 10^{-5}$ with \texttt{BFGS\_bck}, taking 28.23 seconds. Despite this lower runtime, the accuracy gap remained evident.

While analyzing the results under double precision, we observed an interesting phenomenon: certain model-optimizer combinations achieved lower mean absolute error (MAE) on the missing component $f(t)$, despite having a higher final loss. For example, as shown in Table~\ref{tab:PINNs_vs_PIKANs} and Figure~\ref{fig:double_9}, the tanh-cPIKAN model trained with \texttt{BFGS\_bck} reached the lowest MAE of $5.92 \times 10^{-6}$, even though its final loss value was higher than that of the \texttt{BFGS\_trust} configuration, which had a lower loss but a slightly worse MAE of $6.23 \times 10^{-6}$. A similar trend is seen in the PINNs model.

This discrepancy can be attributed to the multi-component nature of the total loss used during training in Phyiscs-Informed Networks. The loss function typically includes multiple terms—data loss, physics-informed residuals, and possibly boundary or regularization constraints—each of which may contribute differently to the total loss. The MAE, on the other hand, reflects the model’s ability to reconstruct the missing function $f(t)$, which is only a subset of the total objective. As such, a model may prioritize optimizing terms other than the one directly tied to $f(t)$, leading to better accuracy (lower MAE) even when the overall loss is higher.

Moreover, second-order optimizers such as \texttt{BFGS\_bck} may refine specific directions in the parameter space that are more relevant to accurately reconstructing $f(t)$, without necessarily minimizing all components of the composite loss. These findings highlight the importance of evaluating models not only based on total loss, but also on task-specific performance metrics—especially in gray-box settings where different loss components may compete during training.

These results highlight the greater benefit of double precision for architectures like tanh-cPIKANs, which involve more complex and sensitive internal representations—such as Chebyshev polynomials. These components rely on fine-grained gradient updates and curvature information, which are better captured under higher precision. Moreover, second-order optimizers—especially BFGS with  trustregion and backtracking linesearch method—depend heavily on accurate gradient and Hessian approximations. Their stopping criteria are directly tied to changes in gradient norms, making them highly responsive to the numerical precision of the training environment. As a result, second-order methods tend to achieve more reliable and lower-error convergence in double-precision settings.

These findings support our motivation for conducting a dedicated study on second-order and hybrid optimizers under both precision regimes. While PINNs offer robustness and consistency across a range of settings, tanh-cPIKANs coupled with appropriate optimization strategies can deliver superior accuracy—particularly when double precision is available.

\subsection{Pharmacodynamics Model}

In this experiment, our objective is to recover the unknown ``chemotherapy efficacy function'' $F_D(t)$ from simulated observations of cancer cell counts $N(t)$, as defined by Eq.~\ref{eq:Pd}. The data consists of $N(t)$ measured every 5 hours over a 600-hour window, yielding discrete time points $t_{\text{data}} = (t_1, t_2, \dots, t_n)$. The function $F_D(t)$ is inherently nonlinear, since the governing pharmacodynamic model in Eq.~\ref{eq:Pd} is a nonlinear ODE. This makes the problem suitable for inverse modeling using neural networks, particularly PINNs and tanh-cPIKANs.

Both models were constrained to have approximately the same number of trainable parameters. After a hyperparameter tuning phase, the architecture for PINNs was chosen as a 5-layer network with 20 neurons per layer, resulting in 1762 trainable parameters. For the tanh-cPIKAN model, we used a 2-hidden-layer architecture with 20 nodes per layer and a Chebyshev polynomial degree of 3, resulting in 1720 trainable parameters.
The number of collocation points for the physics residuals is set to 600.

\subsubsection{Comparison of Model Performance Across First-Order Optimizers}

In this part, we evaluated the performance of best first-order optimizers as shown for previous model, e.g. Adam and RAdam for both PINNs and tanh-cPIKANs models for the pharmacodynamics model.

Both models were trained for the same number of iterations—80,000 epochs—under identical training conditions, including the use of single-precision arithmetic and a cosine learning rate scheduler with an initial learning rate of 0.01.
Table~\ref{tab:cos_pd} summarizes the performance of the models in terms of MAE and computational time across various optimization strategies.

\begin{table}[!h]
\centering
\begin{tabular}{|c|c|c|c|}
\hline
\textbf{Model} & \textbf{Optimizer(\#itr.)} & \textbf{Error (MAE)} & \textbf{Comp. Time} \\
\hline
tanh-cPIKANs & Adam(80k)  & 8.89e-05 & 173.12 \\
tanh-cPIKANs &  RAdam(80k)  & 6.06e-05 & 178.60 \\
\hline
PINNs  & Adam(80k) &  1.09e-04  & 104.98 \\
PINNs  & RAdam(80k) & 9.15e-05 & 105.19 \\

\hline
\end{tabular}
\caption{PD model: Comparison of PINNs and tanh-cPIKANs using different first-order optimizers, which we have shown to perform best in terms of MAE for the gray-box discovery problem. All experiments were performed in single precision, using a cosine learning rate scheduler with an initial learning rate of 0.01.}
\label{tab:cos_pd}
\end{table}

The results indicate that tanh-cPIKANs consistently outperform PINNs in terms of mean absolute error (MAE), achieving a minimum error of \(6.06 \times 10^{-5}\) with RAdam, compared to \(9.15 \times 10^{-5}\) for the best-performing PINNs. While tanh-cPIKANs exhibit higher computational time due to their more expressive adaptive architecture, their improved accuracy suggests a better capacity for learning the nonlinear chemotherapy efficacy function \(F_D(t)\). Notably, RAdam outperforms Adam across both model classes, highlighting its robustness in stabilizing training dynamics, particularly in more sensitive gray-box settings. Under the same training conditions—including the use of RAdam with a cosine learning rate scheduler, identical initial learning rate, nearly equal number of parameters, and the same number of training iterations—both models demonstrated very similar performance, suggesting that in single precision, PINNs and tanh-cPIKANs can be comparably effective when paired with a well-tuned first-order optimizer.

\subsubsection{Performance of PINNs and tanh-cPIKANs with Single Precision}

\begin{figure}[!h] 
    \centering
    \includegraphics[width=0.8\textwidth]{  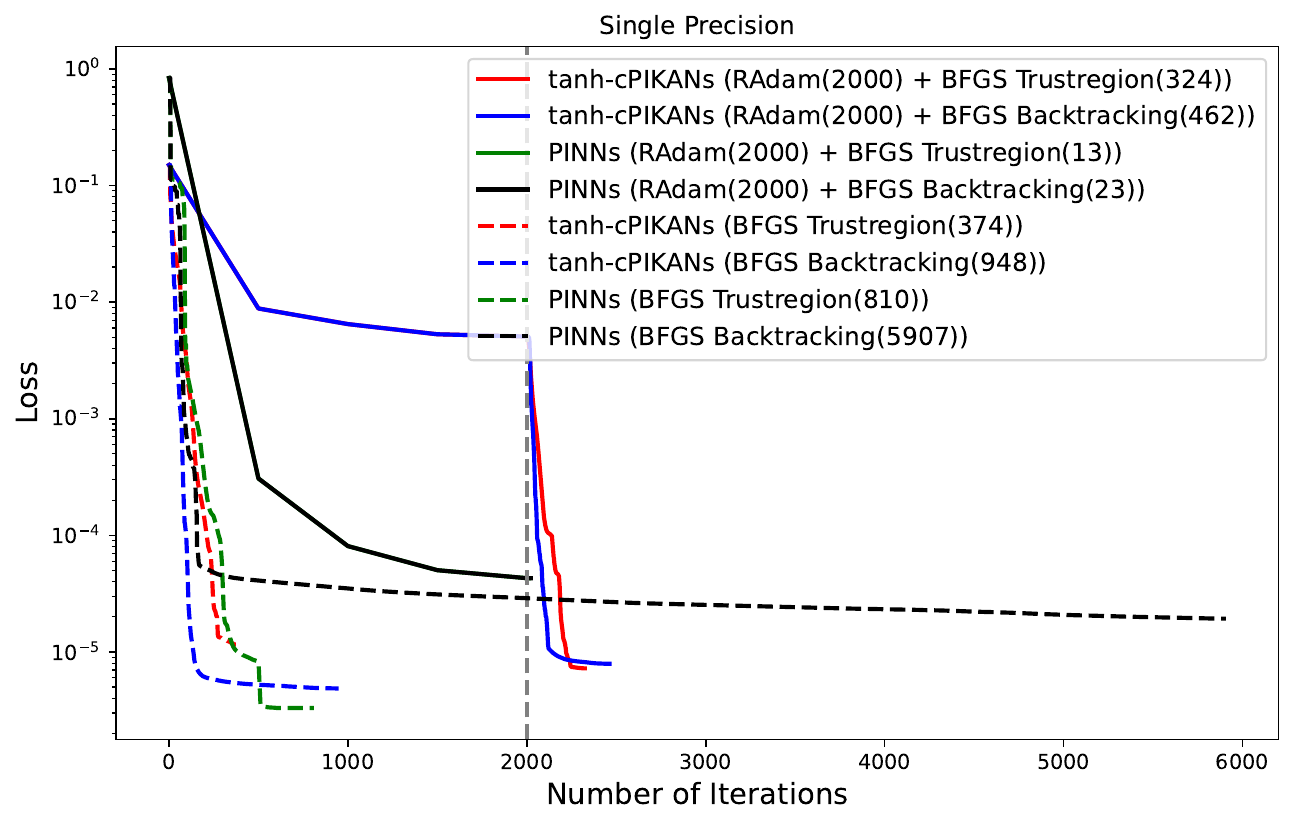} 
    \caption{PD model: Comparison of PINNs and PIKANs performance with different optimizers for single precision.}
    \label{fig:single-8_pd} 
\end{figure}

Table~\ref{tab:PINNs_vs_PIKANs_single_precision_pd} and Figure~\ref{fig:single-8_pd} show the results for the pharmacodynamics (PD) model under single-precision computation, following the same protocol as the PK model. We evaluated second-order optimizers (BFGS with backtracking and trust-region strategies) and hybrid schemes that warm up with RAdam before switching to a second-order optimizer. Based on empirical tuning, we used 2,000 iterations of RAdam before the switch, which proved effective in stabilizing training across both model types.

\begin{table}[!h]
\centering
\begin{tabular}{|c|c|c|c|}
\hline
\textbf{Model} & \textbf{Optimizer} & \textbf{Error (MAE)} & \textbf{Comp. Time} \\
\hline
tanh-cPIKANs & BFGS\_bck(948) & 2.95e-03 & 5.55 \\
tanh-cPIKANs & BFGS\_trust (374) & 5.13e-03 & 5.11\\
tanh-cPIKANs & RAdam(2k) + BFGS\_bck(462) & 2.83e-03 & 17.45 \\
tanh-cPIKANs & RAdam(2k) + BFGS\_trust(324) & 3.57e-03 & 15.70 \\
\hline
PINNs & BFGS\_bck(5.907k) &  3.51e-03 & 16.50 \\
PINNs & BFGS\_trust (810) & 2.68e-03 & 10.05 \\
PINNs & RAdam(2k) + BFGS\_bck(23) & 1.66e-03  & 10.90 \\
PINNs & RAdam(2k) + BFGS\_trust(13) &1.66e-03  & 21.90 \\
\hline
\end{tabular}
\caption{PD model: Comparison of PINNs and tanh-cPIKANs using different optimizers (hybrid or second-order) in terms of MAE and computational time. All experiments were performed in single precision, using a cosine learning rate scheduler with an initial learning rate of 0.001. “BFGS-bck” refers to the BFGS optimizer with a backtracking line search method, while “BFGS-trust” refers to the BFGS optimizer using a trust-region line search method.}
\label{tab:PINNs_vs_PIKANs_single_precision_pd}
\end{table}

The results show that PINNs consistently achieved lower error and faster convergence than tanh-cPIKANs in this single-precision setting. The best-performing PINN achieved a MAE of \(1.66 \times 10^{-3}\) with both RAdam+BFGS-backtracking and RAdam+BFGS-trust, while the best tanh-cPIKAN result was slightly higher at \(2.83 \times 10^{-3}\). Furthermore, PINNs converged significantly faster, with as few as 13 second-order iterations needed in the trust-region case, as opposed to over 300–900 iterations for tanh-cPIKANs.

Figure~\ref{fig:single-8_pd} reinforces these observations. It illustrates a clear distinction in convergence behavior: PINNs exhibit sharp, rapid loss reduction following the optimizer switch, whereas tanh-cPIKANs converge more slowly and to higher loss values. The steep drops in the PINNs loss curves post-warm-up phase highlight how well the hybrid strategy leverages second-order information when gradients are stabilized. In contrast, tanh-cPIKANs appear to plateau earlier, likely due to accumulated precision-related noise impairing second-order curvature estimation.

However, when compared to models trained purely with first-order optimizers over longer durations (e.g., 80k iterations in Table~\ref{tab:cos_pd}), these hybrid and second-order strategies fall short in terms of final accuracy. This suggests that under single precision, second-order optimizers may \textit{prematurely converge} to \textbf{suboptimal local minima}, failing to escape shallow basins due to noisy gradient or curvature signals. The early termination triggered by gradient norm thresholds can cause the optimizer to stop well before reaching a globally optimal solution. Thus, while second-order methods offer rapid convergence, their reliance on precise gradient information makes them less reliable in lower-precision environments—especially for highly nonlinear inverse problems like the PD model.

Overall, these findings emphasize that first-order optimizers, despite their slower progress, can sometimes reach better minima given enough iterations, while second-order methods excel in speed but may require double precision to realize their full potential. A comprehensive set of ablation studies examining the effects of architectural design and training strategies for cPIKAN, tanh-cPIKAN, and PINNs, as well as the impact of adaptive loss weighting for PINNs, is provided in~\nameref{appendix1}.


\subsubsection{Performance of PINNs and tanh-cPIKANs with Double Precision}

\begin{figure}[!h] 
    \centering
    \includegraphics[width=0.8\textwidth]{ 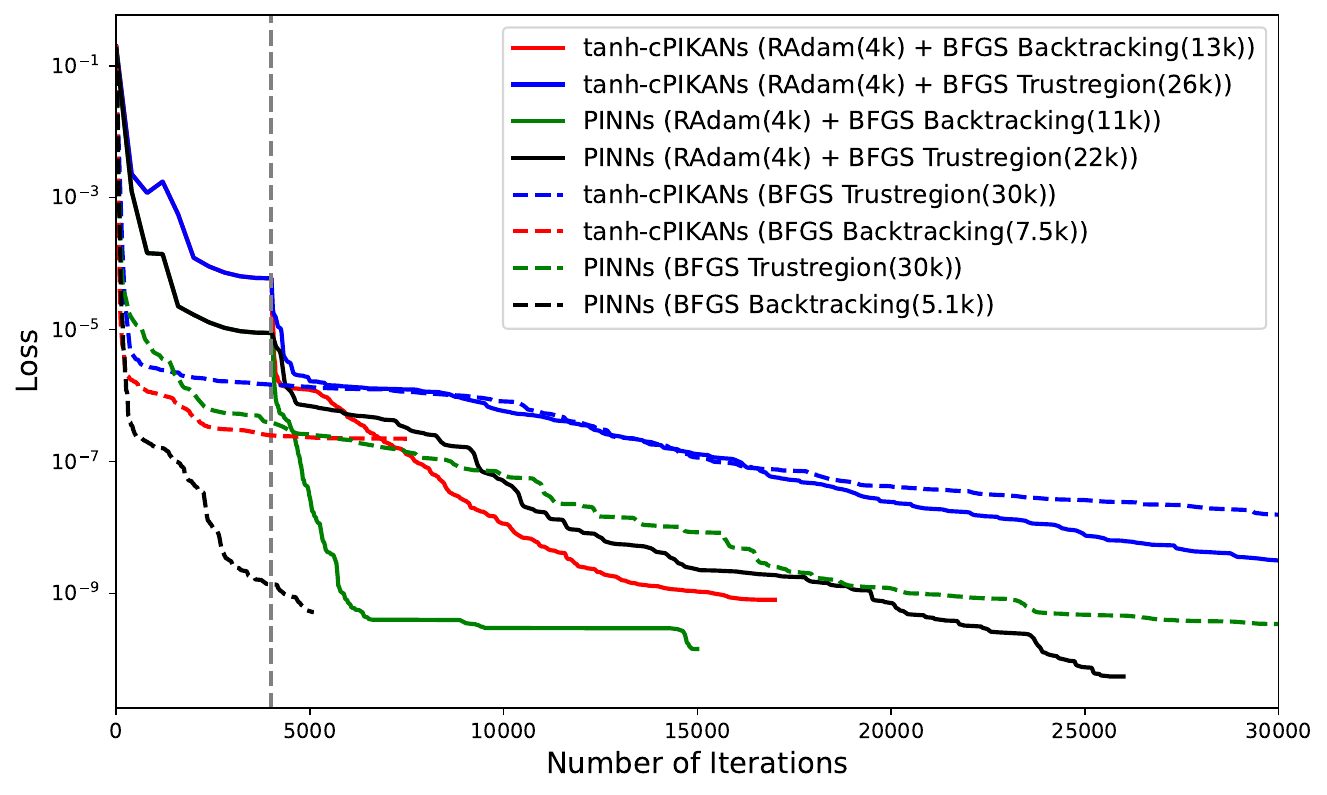} 
    \caption{PD model: Comparison of PINNs and tanh-cPIKANs performance with different optimizers for double precision.}
    \label{fig:double_9_pd} 
\end{figure}

We comprehensively evaluated the performance of PINNs and tanh-cPIKANs under single-precision computation on the pharmacodynamics (PD) model, using various optimization strategies. Table~\ref{tab:PINNs_vs_PIKANs_pd} summarizes the results for second-order and hybrid optimizers, while Figure~\ref{fig:double_9_pd} visualizes the loss trajectories over the course of training.

Both models were trained using BFGS optimizers with either backtracking or trust-region line search, and in hybrid configurations that warm up with 4,000 iterations of RAdam followed by second-order optimization. All experiments used a cosine learning rate scheduler with an initial learning rate of 0.01. For comparison, we also consider results from first-order-only training (80k iterations) shown previously in Table~\ref{tab:cos_pd}.

\begin{table}[!h]
\centering
\begin{tabular}{|c|c|c|c|}
\hline
\textbf{Model} & \textbf{Optimizer} & \textbf{Error (MAE)} & \textbf{Comp. Time} \\
\hline
tanh-cPIKANs & BFGS\_bck(7.5k) & 1.92e-03 & 94.70\\
tanh-cPIKANs & BFGS\_trust (30k) & 2.97e-04 & 373.47\\
tanh-cPIKANs & RAdam(4k) + BFGS\_bck(13k) & \textbf{2.37e-05} & 173.92\\
tanh-cPIKANs & RAdam(4k) + BFGS\_trust(26k) & 7.00e-05 &373.59 \\
\hline
PINNs & BFGS\_bck(5.1k) & 7.67e-05 & 74.63 \\
PINNs & BFGS\_trust (30k) & 8.81e-05 & 474.23 \\
PINNs & RAdam(4k) + BFGS\_bck(11k) & \textbf{4.28e-05} & 188.08 \\
PINNs & RAdam(4k) + BFGS\_trust(22k) & 4.78e-05 & 286.58\\

\hline
\end{tabular}
\caption{PD model: Comparison of PINNs and tanh-cPIKANs using different optimizers (hybrid or second-order) in terms of MAE and computational time. All experiments were performed in double precision, using a cosine learning rate scheduler for RAdam optimizer with an initial learning rate of 0.01.}
\label{tab:PINNs_vs_PIKANs_pd}
\end{table}

The results reveal several notable patterns. First, \textit{hybrid optimization consistently yields the best overall performance for both models}. For tanh-cPIKANs, the lowest MAE of \(2.37 \times 10^{-5}\) was achieved using RAdam followed by BFGS-backtracking, outperforming both pure second-order and pure first-order runs. Similarly, the best PINNs result was \(4.28 \times 10^{-5}\) under the same hybrid scheme.

Second, tanh-cPIKANs exhibit a clear advantage in final accuracy over PINNs when trained in double precision with appropriately tuned hybrid strategies. This is in contrast to the single-precision case (Table~\ref{tab:PINNs_vs_PIKANs_single_precision_pd}), where PINNs were more robust. The improvement in tanh-cPIKANs model performance under double precision suggests that their polynomial-based internal representations, such as Chebyshev expansions, benefit significantly from increased numerical precision—enabling finer gradient updates and more stable second-order optimization.

Figure~\ref{fig:double_9_pd} supports these findings, showing that hybrid training curves (particularly for tanh-cPIKANs) descend sharply and achieve lower loss values. In contrast, purely second-order runs tend to flatten out early, indicating premature convergence. This stagnation is likely due to local minima or saddle points, which second-order methods fail to escape without sufficient warm-up or regularization. Notably, trust-region variants show slower but smoother convergence, while backtracking variants often descend more aggressively.

Moreover, when compared to the fully first-order runs (e.g., 80k RAdam results in Table~\ref{tab:cos_pd}), these hybrid-trained models reach superior or comparable accuracy in significantly fewer total iterations. This illustrates the strength of combining the broad exploration of first-order optimizers with the rapid convergence properties of second-order methods—especially when numerical stability is ensured via double precision.

In summary, these results underscore three main takeaways: (1) hybrid strategies are highly effective, (2) tanh-cPIKANs benefit more from double precision than PINNs, and (3) second-order optimizers alone may prematurely converge unless carefully initialized. These insights reinforce the need to tailor optimization strategies based on architecture complexity and hardware precision.

\subsection{Comparison against Self-Scaled BFGS}

To further assess optimization performance, we compared the standard BFGS and Self-Scaled BFGS (SSBFGS) both for the pharmacokinetics and pharmacodynamics models using the tanh-cPIKANs and PINNs architectures. In this section, we selected the best-performing optimizer settings for each task—whether requiring a warmup phase with RAdam or not—and performed head-to-head comparisons between BFGS and its self-scaled counterpart using the same line search method (either backtracking or trust-region), all under a double-precision setup.

\begin{figure}[!ht] 
    \centering
    \includegraphics[width=0.8\textwidth]{  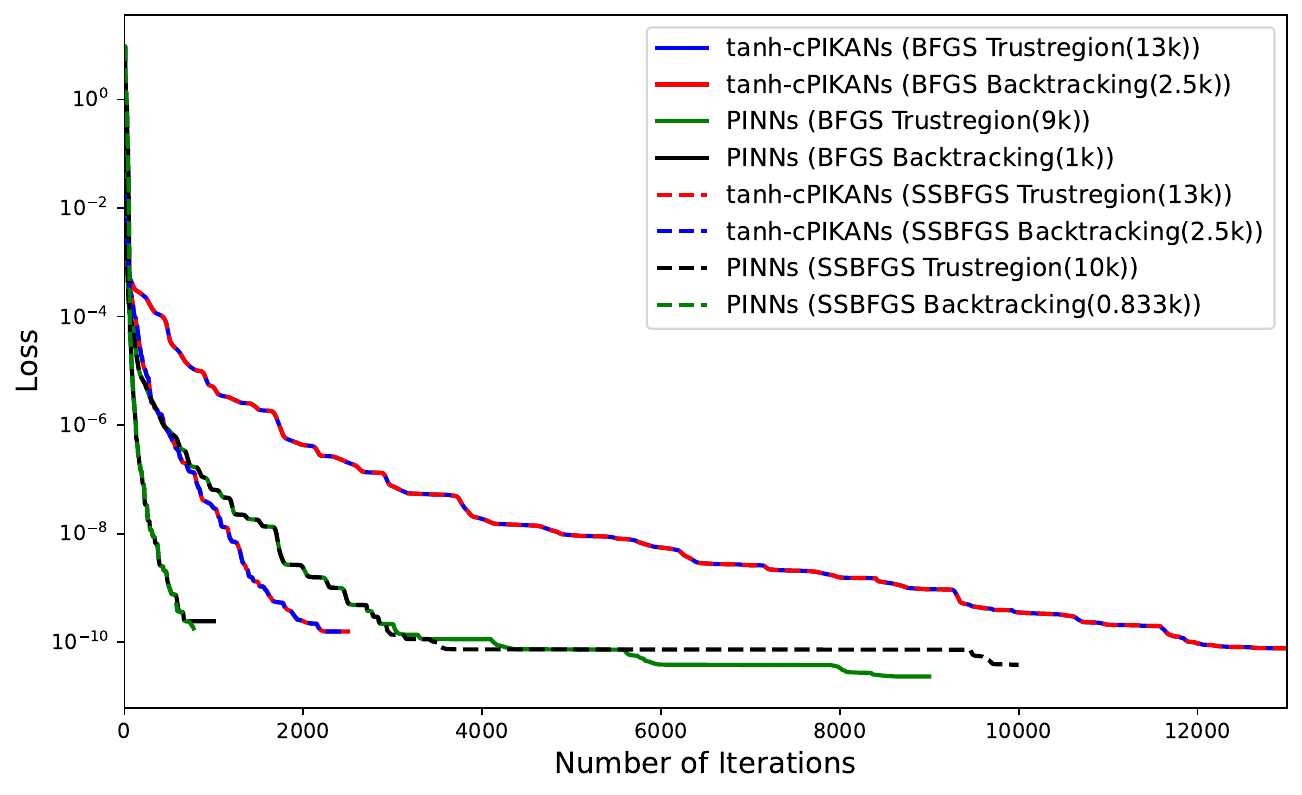} 
    \caption{PK model: Comparison of PINNs and tanh-cPIKANs performance using BFGS and Self-Scaled BFGS (SSBFGS) optimizers with identical line search methods under double precision.}

    \label{fig:double_9_ss} 
\end{figure}

As shown in Table~\ref{tab:PINNs_vs_PIKANs_ssbfgs} and Fig.~\ref{fig:double_9_ss}, tanh-cPIKANs achieved the lowest mean absolute error of \(5.92 \times 10^{-6}\) using either \texttt{BFGS\_bck} or \texttt{SSBFGS\_bck} with 2.5k iterations, requiring approximately 43 seconds of computation. In contrast, PINNs using \texttt{SSBFGS\_bck} reached a slightly higher MAE of \(1.42 \times 10^{-5}\), but did so with significantly lower computation time (7.90 seconds), highlighting their relative efficiency in simpler optimization landscapes.
However, it is important to note that tanh-cPIKANs for the PK model had nearly \textit{four times} the number of trainable parameters (10,300) compared to PINNs (2,752). This larger parameter space introduces more diverse gradient scales and curvature magnitudes across the network, which may diminish the benefit of the global self-scaling mechanism used in SSBFGS. Despite this, tanh-cPIKANs remain more accurate, revealing their architectural strength and robustness. The smoother convergence behavior observed in Fig.~\ref{fig:double_9_ss} further supports the idea that tanh-cPIKANs induce a well-conditioned loss surface that is naturally easier to optimize, even without adaptive weighting.

\begin{figure}[!ht] 
    \centering
    \includegraphics[width=0.8\textwidth]{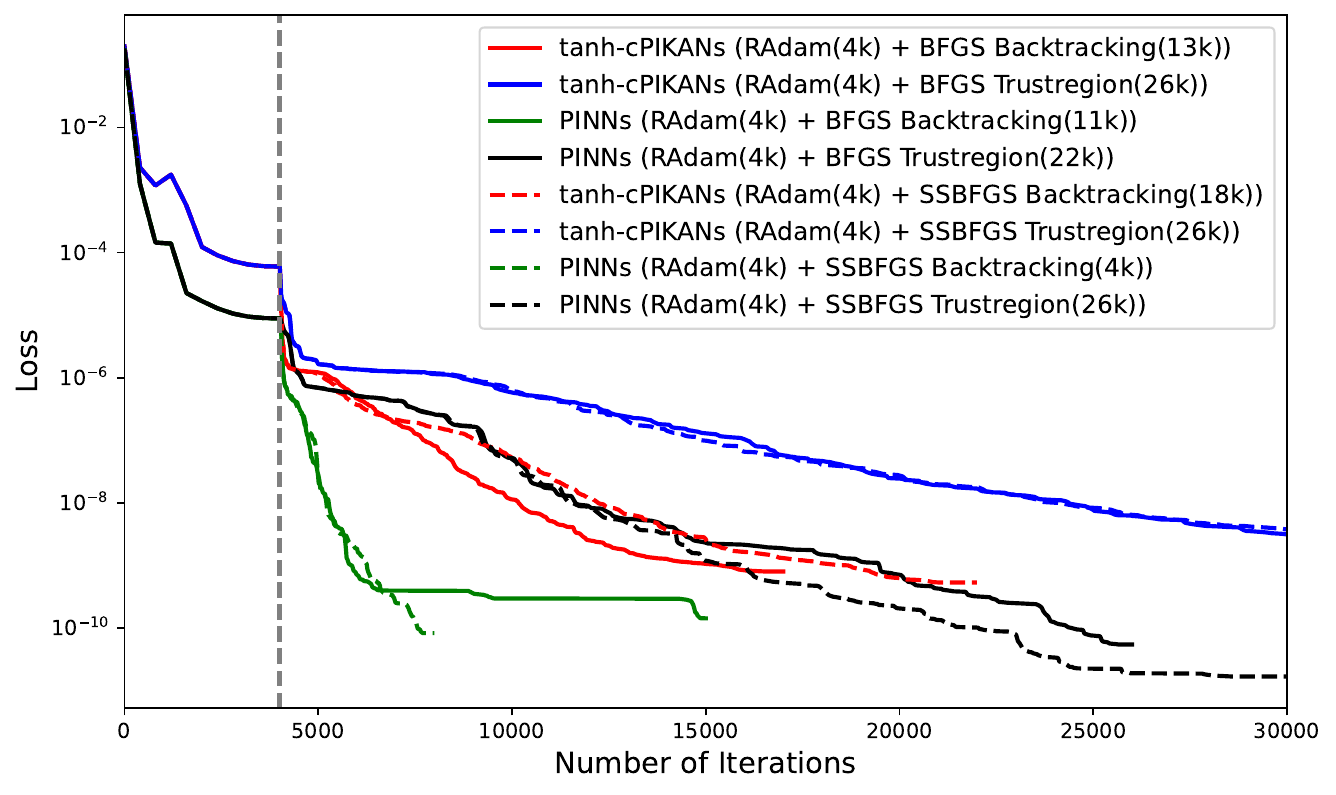} 
    \caption{PD model: Comparison of PINNs and tanh-cPIKANs performance using BFGS and Self-Scaled BFGS (SSBFGS) optimizers with identical line search methods under double precision. Each configuration reflects the best-performing setting for the respective architecture, incorporating warmup phases where beneficial.}

    \label{fig:double_9_pd_ss} 
\end{figure}

In the PD model (Table~\ref{tab:PINNs_vs_PIKANs_pd_ssbfgs}, Fig.~\ref{fig:double_9_pd_ss}), where both models have a comparable number of parameters (tanh-cPIKANs: 1,720 vs. PINNs: 1,762), we observe a more balanced scenario. Here, hybrid training schedules beginning with RAdam and transitioning to a second-order optimizer proved beneficial for both models. Speifically, tanh-cPIKANs achieved the best MAE of \(2.37 \times 10^{-5}\) using \texttt{RAdam(4k) + BFGS\_bck(13k)}, albeit with a higher computational cost of 173.92 seconds. On the other hand, PINNs reached their best performance with \texttt{RAdam(4k) + SSBFGS\_bck(4k)}, attaining a slightly higher MAE of \(3.55 \times 10^{-5}\) in significantly less time—just 45.86 seconds.

These results suggest that the degree of effectiveness of SSBFGS is both architecture- and scale-dependent. While PINNs benefit significantly from self-scaling—especially in lower-parameter or less-structured settings—tanh-cPIKANs, particularly in higher-dimensional regimes, achieve robust performance without relying on such mechanisms. The introduction of the \texttt{tanh} nonlinearity in tanh-cPIKANs helps regularize gradients and stabilize training, while the structured Chebyshev-based representation supports expressive modeling. Together, these properties reduce the need for curvature-aware scaling, making standard BFGS sufficient for efficient convergence.

In summary, self-scaled BFGS often achieves comparable or even better results than standard BFGS across most cases, while maintaining similar or significantly lower computational cost. This makes SSBFGS a competitive default choice for both PINNs and tanh-cPIKANs, particularly in scenarios where training efficiency is critical. For more expressive and higher-capacity networks like tanh-cPIKANs, however, the architectural advantages—such as stability from the \texttt{tanh} nonlinearity and structured Chebyshev representations—reduce the dependency on advanced curvature scaling. As a result, while SSBFGS remains a strong option, standard BFGS is often sufficient to achieve high accuracy, especially in well-regularized training regimes.

\begin{table}[!ht]
\centering
\begin{tabular}{|c|c|c|c|}
\hline
\textbf{Model} & \textbf{Optimizer} & \textbf{Error (MAE)} & \textbf{Comp. Time} \\
\hline
tanh-cPIKANs & BFGS\_bck(2.5K) & \textbf{5.92e-06} & \textbf{43.19} \\
tanh-cPIKANs & BFGS\_trust (13k) & 6.23e-06 & 244.14 \\
tanh-cPIKANs & SSBFGS\_bck(2.5k) & 5.92e-06 & 44.2 \\
tanh-cPIKANs & SSBFGS\_trust(13k) & 6.23e-06 & 244.45 \\
\hline
PINNs & BFGS\_bck(1K) & 2.26e-05 & 28.23 \\
PINNs & BFGS\_trust (9k) & 3.06e-05 & 218.53 \\
PINNs & SSBFGS\_bck(0.833K) & \textbf{1.42e-05} & \textbf{7.90} \\
PINNs & SSBFGS\_trust(10k) & 3.78e-05 & 46.55 \\
\hline
\end{tabular}
\caption{PK model: Comparison of PINNs and tanh-cPIKANs using different optimizers (BFGS and SSBFGS) in terms of MAE and computational time. All experiments were performed in double precision, using a cosine learning rate scheduler for RAdam optimizer with an initial learning rate of 0.001. “BFGS-bck” refers to the BFGS optimizer with a backtracking line search method, while “BFGS-trust” refers to the BFGS optimizer using a trust-region line search method.}
\label{tab:PINNs_vs_PIKANs_ssbfgs}
\end{table}

\begin{table}[!ht]
\centering
\begin{tabular}{|c|c|c|c|}
\hline
\textbf{Model} & \textbf{Optimizer} & \textbf{Error (MAE)} & \textbf{Comp. Time} \\
\hline
tanh-cPIKANs & RAdam(4k) + BFGS\_bck(13k) & \textbf{2.37e-05} & 173.92\\
tanh-cPIKANs & RAdam(4k) + BFGS\_trust(26k) & 7.00e-05 &373.59 \\
tanh-cPIKANs & RAdam(4k) + SSBFGS\_bck(18k) & 4.01e-05  &78.90 \\
tanh-cPIKANs & RAdam(4k) + SSBFGS\_trust(26k) & 6.37e-05 & 120.02\\
\hline
PINNs & RAdam(4k) + BFGS\_bck(11k) & 4.28e-05 & 188.08 \\
PINNs & RAdam(4k) + BFGS\_trust(22k) & 4.78e-05 & 286.58\\
PINNs & RAdam(4k) + SSBFGS\_bck(4k) & \textbf{3.55e-05} & \textbf{45.86} \\
PINNs & RAdam(4k) + SSBFGS\_trust(26k) &  5.06e-05 & 184.25\\

\hline
\end{tabular}
\caption{PD model: Comparison of PINNs and tanh-cPIKANs using different optimizers (BFGS or SSBFGS) in terms of MAE and computational time. All experiments were performed in double precision, using a cosine learning rate scheduler for RAdam optimizer with an initial learning rate of 0.01.}
\label{tab:PINNs_vs_PIKANs_pd_ssbfgs}
\end{table}

\subsection{Robustness of PINNs and tanh-cPIKANs under noisy observations}

In this section, we investigate the robustness of PINNs and tanh-cPIKANs under varying levels of observational noise. Specifically, we examine how each model responds to 5\% and 10\% input noise. We focus on the impact of two second-order optimization methods, BFGS and SSBFGS, combined with two line search strategies: backtracking and trust-region. Our objective is to determine whether architectural choices and optimizer configurations influence the models' ability to generalize from noisy data.

To simulate noisy measurements, we apply multiplicative uniform noise to the sampled outputs from the underlying ODE model. Let \( X(t_i) \in \{G(t_i), B(t_i), U(t_i)\} \) denote the noise-free simulated values at time point \( t_i \), and let \( \eta(t_i) \sim \mathcal{U}(-1, 1) \) be independent uniform random variables. Then, the noisy observations \( \tilde{X}(t_i) \) are generated as:
\begin{equation}
    \tilde{X}(t_i) = X(t_i) \cdot \left(1 + \epsilon \cdot \eta(t_i)\right),
\end{equation}
where \( \epsilon \in \{0.05, 0.10\} \) denotes the noise level. This formulation perturbs each data point independently, introducing variability that mimics real-world measurement noise.\\

To quantify the impact of noise, we report both the absolute MAE values and their relative degradation compared to the noise-free baseline. The degradation factor is computed as the ratio between the MAE under noisy input and the MAE without noise for each method and optimizer configuration:
\begin{equation}
    \text{Degradation Factor} = \frac{\text{MAE}_{\text{noisy}}}{\text{MAE}_{\text{clean}}}.
\end{equation}
This metric helps assess how sensitively each model-optimizer pair responds to increasing levels of noise and highlights configurations that are more robust under uncertainty.

\begin{table}[!ht]
\centering
\begin{tabular}{|c|c|c|c|c|}
\hline
\textbf{Model} & \textbf{Optimizer} & \multicolumn{3}{c|}{\textbf{MAE}} \\
\cline{3-5}
 &  & \textbf{noise 0\%} & \textbf{noise 5\%} & \textbf{noise 10\%} \\
\hline
tanh-cPIKANs & BFGS\_bck & 5.92e-06 & 9.95e-05 (×16.8) & 2.33e-04 (×39.4) \\
tanh-cPIKANs & BFGS\_trust & 6.23e-06 & 9.51e-05 (×15.3) & 1.19e-04 (×19.1) \\
tanh-cPIKANs & SSBFGS\_bck & 5.92e-06 & 9.63e-05 (×16.3) & 2.07e-04 (×35.0) \\
tanh-cPIKANs & SSBFGS\_trust & 6.23e-06 & 6.29e-05 (×10.1) & 1.39e-04 (×22.3) \\

\hline
PINNs & BFGS\_bck     & 2.26e-05 & 6.18e-04 (×27.3) & 6.97e-04 (×30.8) \\
PINNs & BFGS\_trust   & 3.06e-05 & 4.73e-04 (×15.5) & 6.87e-04 (×22.5) \\
PINNs & SSBFGS\_bck   & 1.42e-05 & 6.25e-04 (×44.0) & 8.25e-04 (×58.1) \\
PINNs & SSBFGS\_trust & 3.78e-05 & 7.84e-04 (×20.7) & 9.12e-04 (×24.1) \\

\hline
\end{tabular}
\caption{PK model: Evaluation of the robustness of PINNs and tanh-cPIKANs using different optimizers (BFGS and SSBFGS) in terms of MAE under noisy observation. All experiments were performed in double precision, using a cosine learning rate scheduler for the RAdam optimizer with an initial learning rate of 0.001. “BFGS-bck” refers to the BFGS optimizer with a backtracking line search method, while “BFGS-trust” refers to the BFGS optimizer using a trust-region line search method. Results are reported for two levels of input noise: 10\% and 5\%, applied independently to each method to assess robustness.}
\label{tab:noise}
\end{table}

Table~\ref{tab:noise} presents the MAE results across clean and noisy settings. Overall, tanh-cPIKANs outperform PINNs in both accuracy and robustness to observational noise. At noise levels of $5\%$ and $10\%$, tanh-cPIKANs consistently achieve lower error rates and exhibit more stable relative degradation compared to their noise-free baseline. Notably, tanh-cPIKAN paired with SSBFGS\_trust shows the smallest degradation, increasing only by 10.1× at $5\%$ noise and 22.3× at $10\%$ noise—representing the most stable performance among all configurations. Trust-region variants also improve robustness in other tanh-cPIKAN settings, with degradation remaining under 20× at $5\%$ noise across the board.
In contrast, PINNs demonstrate significantly higher sensitivity to noise, especially with second-order optimizers. PINNs with SSBFGS\_bck show the most severe error amplification, with 44.0× and 58.1× degradation at $5\%$ and $10\%$ noise, respectively. However, for some PINN configurations, the degradation from 5\% to 10\% noise is less drastic, likely because the error is already high at the 5\% level—leaving less room for proportional increase. This saturation effect suggests that PINNs may hit a noise sensitivity ceiling earlier than tanh-cPIKANs.
It is also worth noting that although tanh-cPIKAN with BFGS\_bck maintains low absolute errors, it shows relatively higher degradation at $10\%$ noise (×39.4), likely due to the instability of backtracking-based line searches in high-noise regimes. This emphasizes that while model architecture plays a critical role, the choice of line search strategy can strongly influence robustness.
These results confirm that robustness to observational noise is significantly enhanced by combining expressive architectures like tanh-cPIKANs with second-order optimizers equipped with conservative trust-region strategies. Such configurations offer a reliable pathway for building noise-resilient gray-box models in scientific applications.

\section{Summary and Outlook}
In this work, we systematically investigated the optimization of two different physics-informed networks  for gray-box discovery problems, focusing particularly on inverse problems in pharmacokinetics (PK) and pharmacodynamics (PD). While we used relatively simple PK/PD models and simulated data,  our primary aim was to answer key questions regarding optimizer suitability and training stability— especially in the context of ill-posed, sparse, and non-unique problems that arise frequently in 
PK/PD. To address these challenges, we introduced a slightly new  architecture—\textit{tanh-cPIKANs}—a variant of Chebyshev-based Kolmogorov–Arnold Networks (cPIKANs) with improved convergence and numerical stability. This design enhances gradient smoothness through the use of bounded activations (tanh), making it particularly effective for both first- and second-order optimization techniques (see also \ref{appendix2}).

Using extensive benchmarks, we found that no single optimizer was universally optimal across models and settings. However, hybrid strategies—particularly \textit{RAdam followed by BFGS with either backtracking or trust-region line search}—consistently delivered the best results for both PINNs and tanh-cPIKANs in terms of accuracy and training efficiency. Among first-order methods, RAdam proved to be the most effective and robust, benefiting from adaptive momentum while avoiding some of the instability seen with plain Adam. For learning rate scheduling, the cosine decay consistently improved performance, but its effectiveness was highly dependent on the choice of initial learning rate. We observed that a higher starting learning rate was beneficial when combined with a warm-up phase, especially for architectures with more complex internal representations like cPIKANs.

Our study also demonstrated that \textit{model representation plays a critical role} for enhanced accuracy. While standard MLP-based PINNs offered greater robustness in single precision, tanh-cPIKANs outperformed them in double precision, achieving lower MAE in fewer iterations when paired with properly initialized second-order optimizers. This highlights a key trade-off: expressive models such as cPIKANs can achieve superior accuracy but are more sensitive to precision and optimization. These models, in particular, benefit from the curvature information exploited by second-order methods, which in turn require stable gradients and higher floating-point fidelity.

We further showed that warm-up phases with first-order optimizers provide more robust training for second-order methods to reach optimal performance. Without warm-up, BFGS variants may converge prematurely to local minima due to poor initial curvature estimation—especially under single-precision arithmetic. This sensitivity underscores the importance of precision: \textit{double precision training significantly improves optimizer behavior}, particularly for second-order methods that rely on gradient norms for convergence criteria. {\color{blue} Although our study focuses on low-dimensional ODE-based PK/PD systems, the utilization of second-order and quasi-Newton methods can also be effective in large-scale settings such as PDEs, stiff dynamics, or high-dimensional parameter spaces. The scalability of second-order optimizers (those used in this study) primarily depends on the number of trainable parameters in the neural network. The dimensionality of the underlying PDE does not directly affect scalability because the PDE inputs can be sampled stochastically during each training iteration, keeping the effective computational load independent of the PDE dimension. Furthermore, even when the neural network contains a large number of parameters, the proposed optimizers can remain practical. This is because the optimizer does not require updating every parameter explicitly at each step; instead, it relies on gradient information and low-rank approximations to the Hessian. In addition, the Hessian (or its approximation) can be constructed using gradients computed for a randomly selected subset of parameters, which further reduces computational costs while still capturing sufficient curvature information to guide optimization effectively. As a result, the proposed optimizers can handle relatively high-dimensional parameter spaces more efficiently than full Newton-type methods while still retaining key second-order advantages. A recent complementary study~\cite{elham} shows that self-scaled quasi-Newton methods (e.g., SSBFGS, SSBroyden) remain highly effective for challenging PDE problems, but also identifies practical limitations related to memory growth and line-search sensitivity. We therefore note that while our optimization insights extend conceptually to more complex systems, their efficiency in large-scale regimes depends on memory-aware implementations and problem-specific numerical considerations.}

Across both PK and PD models, we observed that self-scaled BFGS  offered consistently competitive performance compared to standard BFGS. In many cases, SSBFGS achieved comparable or lower errors while requiring equal or reduced computational time, making it a practical and efficient second-order optimization strategy. Notably, for PINNs—especially in lower-dimensional settings like the PD model—SSBFGS often outperformed BFGS in terms of convergence speed. While the benefits were less pronounced for tanh-cPIKANs in the high-parameter PK model, this was likely due to the architecture’s inherent stability and smoother loss landscape, which made the added scaling of SSBFGS less critical. These results suggest that SSBFGS is a reliable and efficient alternative to classical BFGS, particularly when model scale or curvature complexity varies.

While our results provide valuable guidelines, several limitations remain. First, our study focused on moderately sized models and problems; the suitability of second-order or quasi-Newton methods for larger-scale PDE-constrained systems still requires careful memory and efficiency considerations. Second, although JAX enabled rapid prototyping and efficient automatic-differentiation, its current limitations in mixed precision and full GPU memory utilization for large Hessian computations suggest further engineering optimizations are needed. Additionally, the performance of other KAN variants (e.g., rational or ReLU-based) was not investigated and is left for future work.

In conclusion, our findings demonstrate that \textit{optimizer choice, model architecture, and numerical precision are tightly coupled} in gray-box discovery problems. The new tanh-cPIKANs represent a promising direction for scientific modeling tasks that demand expressiveness and smooth optimization landscapes. Importantly, our robustness study further reveals that the interplay between architecture and optimization strategy also governs the model's ability to handle noisy data—an essential consideration for real-world applications. The full potential of tanh-cPIKANs is realized when paired with hybrid training strategies, robust second-order optimizers, and sufficient numerical precision, reinforcing the importance of principled design choices in physics-informed learning pipelines.


\textcolor{blue}{\section*{Code and Data Availability}}
\textcolor{blue}{All codes, trained models, and synthetic PK/PD datasets used in this study are publicly available at \url{https://github.com/NazAhmadi/PINNs_PIKANs_Graybox}.}

\vspace{1.9em} 
\section*{Acknowledgements}
This work was supported by the National Institutes of Health (NIH) grant $ R01HL154150$.


\newpage

\bibliographystyle{elsarticle-num} 
\bibliography{main}

\appendix
\newpage

\section*{Appendix}
\phantomsection
\addcontentsline{toc}{section}{Appendix}

\renewcommand{\thesubsection}{A.\arabic{subsection}}
\setcounter{subsection}{0}

\textcolor{blue}{\subsection{Parameter-Matched PK Experiments}}
\label{appendixnew}

\textcolor{blue}{To further examine whether differences in parameter count influence performance in the pharmacokinetics case study, we conducted a series of additional experiments using parameter-matched and reduced-parameter architectures for both tanh-cPIKANs and PINNs. All experiments were performed using a fixed learning rate of 0.001 with Adam or RAdam optimizers. The results clearly show that smaller tanh-cPIKAN architectures (2k–3k parameters) outperform substantially larger PINNs (over 10k parameters), confirming that the performance gap arises from architectural and optimization advantages rather than raw parameter capacity. The full results are summarized in Table~\ref{tab:pk_clean_results}.}
\begin{table}[!h]
\centering
\small
\begin{tabular}{|c|c|c|c|c|}
\hline
\textbf{Model / Basis} & \textbf{Architecture} & \textbf{\#Params} & \textbf{Optimizer} & \textbf{MAE} \\
\hline

PINN  & (50, 2) & 2752 & Adam  & $9.15\times10^{-05}$ \\ \hline
PINN  & (50, 2) & 2752 & RAdam & $7.45\times10^{-05}$ \\ \hline

tanh-cPIKANs & (20,3,3) & 3320 & RAdam & $3.07\times10^{-5}$ \\ \hline
Jacobi PIKANs $(1,1)$ & (20,3,3) & 3320 & RAdam & $6.85\times10^{-5}$ \\ \hline
Jacobi PIKANs $(0,0)$ & (20,3,3) & 3320 & RAdam & $5.25\times10^{-5}$ \\ \hline

tanh-cPIKANs & (20,3,3) & 3320 & Adam  & $5.19\times10^{-5}$ \\ \hline
tanh-cPIKANs & (16,3,3) & 2144 & Adam  & $2.75\times10^{-5}$ \\ \hline
tanh-cPIKANs & (16,3,3) & 2144 & RAdam & $3.19\times10^{-5}$ \\ \hline

PINNs & (50,5) & 10402 & Adam  & $8.62\times10^{-5}$ \\ \hline
PINNs & (50,5) & 10402 & RAdam & $1.18\times10^{-4}$ \\ \hline

tanh-cPIKANs & (50, 2, 3) & 10300 & Adam  & $5.25\times10^{-05}$ \\ \hline
tanh-cPIKANs & (50, 2, 3) & 10300 & RAdam & $8.42\times10^{-05}$ \\ \hline

\end{tabular}

\caption{\textcolor{blue}{Additional PK experiments evaluating parameter-matched and original architectures for Jacobi PIKANs, tanh-cPIKANs, and PINNs. The table includes a direct comparison of Chebyshev-based PIKANs with Jacobi-based PIKANs using the same architecture \((20,3,3)\) and optimizer (RAdam, LR = 0.001). For Jacobi PIKANs, $(\alpha,\beta)$ denotes the Jacobi shape parameters. Jacobi $(1,1)$ produces noticeably higher error, whereas Jacobi $(0,0)$ achieves accuracy comparable to Chebyshev but requires substantially higher computational cost (approximately $4$--$5\times$ slower). Across all configurations, smaller tanh-cPIKANs (2144–3320 parameters) consistently outperform both larger tanh-cPIKANs (10{,}300 parameters) and significantly larger PINNs (10{,}402 parameters), confirming that performance gains stem from architectural and optimization advantages rather than parameter count.}}

\label{tab:pk_clean_results}
\end{table}

\subsection{Ablation Study for Pharmacokinetics Model — PINNs}
\label{appendix1}
In this section, we conduct a series of ablation studies to investigate the impact of key design choices on the performance of PINNs. Unlike the main experiments in the paper, which used PINNs with fewer parameters for fair comparison, here we explore a larger architecture with over 10,000 trainable parameters to assess scalability and flexibility. Specifically, we examine: (i) the effect of using two neural networks to capture potentially missing components or to decouple dynamics; (ii) the influence of adaptive loss weighting compared to fixed weights; and (iii) the benefit of a two-step training strategy, where the model is initially trained using only data loss before introducing physics-based constraints. In the two-step setting, the first training phase is performed solely on data loss, and the second phase incorporates both data and physics-based losses. All experiments were conducted using the Adam optimizer with a cosine learning rate scheduler initialized at 0.001. The results are summarized in table \ref{tab:ablation_results_pinns}.

\begin{table}[ht]
\centering
\begin{tabular}{|c|c|c|c|c|c|}
\hline
\textbf{Case} & \textbf{Training} & \textbf{Adap. W} & \textbf{Architecture} & \textbf{Error (MAE)} & \textbf{Time} \\
\hline
1 & 5k + 45k & Yes & [50, 6] & 3.15e-05 & 375.81 \\
2 & 5k + 45k & No & [50, 6] & 6.78e-05 & 366.80 \\
3 & 50k & Yes & [50, 6] & \textbf{2.16e-05} & 373.67 \\
4 & 50k & No & [50, 6] & 4.02e-05 & 350.60 \\
5 & 5k + 45k & No & [50, 6], [20, 4] & 2.24e-05 & 378.25 \\
6 & 50k & No & [50, 6], [20, 4] & 6.36e-05 & 376.21 \\
7 & 5k + 45k & Yes & [50, 6], [20, 4] & 2.49e-05 & 379.29 \\
8 & 50k & Yes & [50, 6], [20, 4] & 4.10e-05 & 376.39 \\
\hline
\end{tabular}
\caption{Ablation study results for PINNs under different training strategies, network configurations, and loss weighting schemes. All models were trained using the Adam optimizer with a cosine learning rate scheduler (initial learning rate = 0.001). The "Architecture" column indicates the number of neurons and hidden layers; if a second bracketed architecture is present, it corresponds to an auxiliary network used to identify missing components in the governing equations.}
\label{tab:ablation_results_pinns}
\end{table}

\subsection{Ablation Study for Pharmacokinetics Model — PIKANs}
\label{appendix2}

We conducted an ablation study on the cPIKAN and tanh-cPIKAN models to evaluate the effects of polynomial degree, outer nonlinearity, and training strategy for the pharmacokinetics model. All models were trained for 70,000 iterations using the RAdam optimizer with a cosine learning rate scheduler initialized at 0.01. We first compared a two-step training scheme—comprising 5k iterations focused solely on data loss followed by 65k iterations including the physics-informed loss—with a single-step direct training approach. For the same architecture \([50, 2, 1]\), the two-step strategy achieved a slightly improved MAE of \(1.77 \times 10^{-4}\) compared to \(2.16 \times 10^{-4}\) for one-step training, with similar computational costs. Increasing the polynomial degree from 1 to 3 led to a noticeable reduction in error (MAE: \(1.31 \times 10^{-4}\)), suggesting that enhanced function approximation via higher-degree polynomials contributes positively to expressivity. Further improvements were obtained with the tanh-cPIKAN architecture \([50, 2, 3]\), which incorporates an outer tanh nonlinearity and achieved the best overall performance (MAE: \(\mathbf{1.24 \times 10^{-5}}\)) with moderate additional computational time. However, increasing the polynomial degree to 5 resulted in better accuracy for cPIKANs (MAE: \(9.69 \times 10^{-5}\)) and a substantial increase in training cost. These findings underscore the importance of balancing model capacity with optimization efficiency and highlight the advantage of outer nonlinearities in stabilizing and enhancing learning in KAN-based gray-box models. The results are summarized in table 
\ref{tab:ablation_results_cpikan}.

\begin{table}[ht]
\centering
\begin{tabular}{|c|c|c|c|c|c|c|}
\hline
\textbf{Case} & \textbf{Training} & \textbf{Model} & \textbf{Architecture} & \textbf{Error (MAE)} & \textbf{Time} \\
\hline
1 & 5k + 65k & cPIKANs & [50, 2, 1] & 1.77e-04 & 141.02 \\
2 & 70k & cPIKANs &  [50, 2, 1] & 2.16e-04 & 138.00 \\
3 & 70k & cPIKANs &  [50, 2, 3] & 1.31e-04 & 255.90 \\
4 & 70k & tanh-cPIKANs &  [50, 2, 3] & \textbf{1.24e-05} & 305.02 \\
5 & 70k & cPIKANs &  [50, 2, 5] & 9.69e-05 & 494.56 \\
\hline
\end{tabular}
\caption{Ablation study results for cPIKAN and tanh-cPIKAN models with varying polynomial degrees and training procedures. All models were trained using RAdam with a cosine scheduler (initial learning rate = 0.01).  In the "Architecture" column, the first number indicates the number of nodes, the second represents the number of hidden layers, and the third denotes the polynomial degree.}
\label{tab:ablation_results_cpikan}
\end{table}

\subsubsection{Ablation Study on Architecture Design for cPIKAN and tanh-cPIKAN Models}

We further conducted an ablation study to explore how architectural design choices—such as the polynomial degree and network configuration—affect the performance of tanh-cPIKAN and cPIKAN models. All models were trained for 70,000 iterations using either the Adam or RAdam optimizer with a cosine learning rate scheduler (initial learning rate = 0.01). The architectures are denoted as \([n, l, d]\), representing the number of neurons per layer, number of hidden layers, and polynomial degree, respectively. In this study, two parallel networks were used to model different components of the gray-box formulation. Among the configurations, the tanh-cPIKAN model with architecture \([50,2,3], [20,2,3]\) trained with Adam achieved the best performance, with MAE of \(6.11\times10^{-5}\). Reducing the polynomial degree or altering the hidden layers of the second network slightly degraded accuracy, while switching to RAdam significantly worsened performance for both tanh-cPIKAN and standard cPIKAN models. These results highlight that model expressivity and optimization stability are highly sensitive to architectural configurations and optimizer choice. 
The results are summarized in table 
\ref{tab:ablation_architecture}.
\begin{table}[!ht]
\centering
\begin{tabular}{|c|c|c|c|c|c|}
\hline
\textbf{Training} & \textbf{Optimizer} & \textbf{Model} & \textbf{Architecture} & \textbf{Error (MAE)} & \textbf{Time} \\
\hline
70k & Adam & tanh-cPIKANs & [50,2,3], [20,2,3] & 6.11e-05 & 411.71 \\
\hline
70k & Adam & tanh-cPIKANs & [50,2,1], [20,2,1] & 9.38e-05 & 269.45 \\
\hline
70k & Adam & tanh-cPIKANs & [50,2,3], [20,3,1] & 1.14e-04 & 421.55 \\
\hline
70k & RAdam & tanh-cPIKANs & [50,2,3], [20,2,3] & 2.65e-03 & 414.64 \\
\hline
70k & RAdam & cPIKANs & [50,2,3], [20,2,3] & 2.71e-03 & 397.16 \\
\hline
\end{tabular}
\caption{Ablation study on architectural variations for cPIKAN and tanh-cPIKAN. Two-network configurations are evaluated across different depths, polynomial degrees, and optimizers. All models were trained using a cosine scheduler (initial learning rate = 0.01) for 70k iterations.}
\label{tab:ablation_architecture}
\end{table}

\end{document}